\documentstyle[12pt]{article}
\title{The phase-space structure of \\the Klein--Gordon field}
\author{
  Christoph Best$^{1,2}$, Pawe{\l} G\'ornicki$^{3,4}$, and
  Walter Greiner$^1$
  \\[2mm]
  {\small\sl $^1$Institut f\"ur Theoretische Physik,
                 Johann Wolfgang Goethe-Universit\"at,} \\
  {\small\sl 6000 Frankfurt am Main, Germany,} \\
  {\small\sl $^2$School of Physics and Astronomy,
                 Raymond and Beverly Sackler Faculty } \\
  {\small\sl of Exact Sciences, Tel Aviv University, 69978 Tel Aviv, Israel,}
\\
  {\small\sl $^3$Max-Planck-Institut f\"ur Kernphysik,} \\
  {\small\sl Postfach 10 39 80, D-6900 Heidelberg, Germany,} \\
  {\small\sl $^4$Centrum Fizyki Teoretycznej,}\\
  {\small\sl Al. Lotnik\'ow 32/46, PL-02-668 Warsaw, Poland.}
}
\date{January 27, 1993}

\begin{document}
\maketitle

\begin{abstract}
The formalism based on the equal-time Wigner function of the two-point
correlation function for a quantized Klein--Gordon field is presented. The
notion of the gauge-invariant Wigner transform is introduced and equations
for the corresponding phase-space calculus are formulated. The equations of
motion governing the Wigner function of the Klein--Gordon field are derived.
It is shown that they lead to a relativistic transport equation with
electric and magnetic forces and quantum corrections. The governing
equations are much simpler than in the fermionic case which has been treated
earlier. In addition the newly developed formalism is applied towards the
description of spontaneous symmetry breakdown.
\end{abstract}

\section{Introduction}                                    \label{secIntro}

We consider the time evolution of quantum fields interacting with a
classical (external) field. A useful example of such a system is the
particle creation by strong but slowly varying electromagnetic fields. In
this case the particle fields require full quantum treatment while the
electromagnetic field may be considered classically. It is sometimes
necessary to treat these classical fields as fully dynamical objects rather
than as a fixed  background. The study of the back reaction problem is such
a case. Quite recently this type of dynamical description has been
successfully applied \cite{BGR} towards the description of the
electron--positron vacuum in QED. Eventually some of these approachs may
serve as crude models for the formation stage of the quark-gluon plasma.

To achieve their goals the authors of \cite{BGR} had to simplify the notion
of the vacuum by truncating the infinite hierarchy of correlation functions.
While the full description of a vacuum state at a given instant of time is
provided by the set of all possible $n$-point equal-time correlation
functions of the quantum field, two-point functions are sufficient to
construct the electric current and therefore to describe the influence of
the electron--positron vacuum on a classical electromagnetic field. The
authors of \cite{BGR} make use of the Fourier-transformed two-point
equal-time correlation function (an analog of the Wigner function) for
fermionic fields. The resulting equations describe the evolution of the
vacuum from a phase-space point of view. Our goal here is to extend the
basic results presented in the cited work to the case of scalar
electrodynamics. The relative simplicity of the scalar result helps in the
interpretation of the physical features of the interacting system.

The use of the Wigner functions to the description of the QFT vacuum is not
new \cite{Eisen,Vasak,Elze}. Neither is the use of the two-point correlation
function to describe the interaction with a classical but dynamical
electromagnetic field \cite{CM1,CM2,CM3}. The main difference is that we use
equal-time correlation functions and do not perform the Fourier transform
with respect to the time variable. (The equal-time approach is also used in
Ref.~\cite{Provid}.) The equation of motion is therefore a true evolution
equation that can be numerically solved as an initial-value problem.

In section \ref{secCalc} we introduce the gauge-invariant Wigner transform
and show how the noncommutative operator algebra of quantum mechanics is
implemented in the space of phase-space transforms. The Wigner transform of
the equal-time two-point correlation function of the Klein--Gordon field in
Feshbach-Villars representation is defined and discussed in section
\ref{secWigner}, which will, incidentally, illustrate the power of the
methods developed in section \ref{secCalc}. Finally, in section \ref{secSSB}
we apply this formalism to spontaneous symmetry breakdown.

\section{Wigner function calculus}                         \label{secCalc}
The Wigner transform associates a quasiprobability distribution in
classical phase space to every quantum-mechanical density matrix in such a
way that quantum-mechanical averages can be computed like classical
statistical averages. The equation of motion of the quasiprobability
distribution derives from the quantum-mechanical Hamiltonian and corresponds
to the classical Liouville equation with quantum corrections which
can be given as an expansion in $\hbar$. Since the Liouville equations is
easier to solve than the full quantum treatment, the Wigner function
provides a tool to simplify calculations when semiclassical approximations
are valid.

Similarly, the two-point function of a quantum field theory can be
interpreted as the matrix element of a density matrix. The resulting
phase-space quasiprobability distributions reflect the many-particle
content of the quantum field theory. Their equation of motion follows
analogously from the operator equation of motion, i.e., the wave equation,
and the corresponding Hamiltonian.

Mathematically, the Wigner transform provides a realization of the abstract
Heisenberg operator algebra of quantum mechanics on the space of functions
over phase space, thus translating every operator equation---like the
Heisenberg equation of motion---into a differential-equation on phase
space---like the Liouville equation. Especially, it induces a noncommutative
multiplication on this function space that is characteristic of the
quantum-mechanical content of the theory. Once the representation of the
basic operators and this multiplication is known, every operator equation
can easily and without explicit calculation be translated into a phase-space
equation.

In the following, we discuss this calculus for a Wigner function exhibiting
local U(1)-gauge invariance. The notational conventions are from the review
article \cite{Hil}, where the gauge issue was not considered. We set $c=1$
but keep $\hbar$ in this section since expansions around the classical limit
are frequent.

\subsection{Definition and elementary properties}
Let us consider a $D$-dimensional configuration space of vectors $q\in{\bf
R}^D$ and the corresponding Hilbert space $\cal H$, spanned by basis vectors
$\mbox{$\langle q|$}$. Suppose that the operator $\hat A$ transforms
covariantly under a
local U(1)-gauge transformation $\Lambda(q)$
\begin{equation}
  \mbox{$\langle {q}|{\hat A}|{q'}\rangle$} \to
  e^{ie\{\Lambda(q) - \Lambda(q')\}}
\mbox{$\langle {q}|{\hat A}|{q'}\rangle$}.
\end{equation}
The corresponding transformation of the vector potential ${\cal A}(q)
\in{\bf R}^D$ is
\begin{equation}
  {\cal A}(q) \to {\cal A}(q) + \frac{\partial \Lambda}{\partial q}(q).
\end{equation}
Every operator $\hat A$ may be associated with a function in phase-space
$(q,p)\in{\bf R}^D\times{\bf R}^D$
\begin{equation} \label{Wig31}
  P_{\hat A}(q,p) =
    \int {\rm d}^{D}{y} \,
    \mbox{$\langle {q-\frac{1}{2} y}|{\hat A}|
  {q+\frac{1}{2} y}\rangle$} \,
    \exp\left(\frac{i}{\hbar} p\cdot y + \frac{ie}{\hbar}
              \int_{q-y/2}^{q+y/2} {\cal A}(x)\cdot{\rm d}{x}\right).
\end{equation}
This is an obvious generalization of the well-known Wigner transformation,
and we will use this name here. Indeed, $P_{\hat\rho}(q,p)$ ($\hat\rho$
being a density matrix) is the usual Wigner phase-space distribution.
The line integral in Eq. (\ref{Wig31}) is taken over a straight line, i.e.,
\begin{equation}
  \int_{q-y/2}^{q+y/2} {\cal A}(x)\cdot{\rm d}{x} =
  \int_{-1/2}^{+1/2} {\rm d}{\lambda} y\cdot {\cal A}(q + \lambda y)
\end{equation}
where the dot indicates the scalar product in $D$ dimensions. This factor
makes the resulting function $P_{\hat A}(q,p)$ a gauge-invariant realization
of the abstract operator $\hat A$ in phase space. Every operator equation
can thus be translated into a phase-space equation. In the following we
discuss the required formalism and apply it to the Heisenberg equation of
motion for a density matrix.

The inverse of the transformation (\ref{Wig31}) is
\begin{equation} \label{invWig31}
  \mbox{$\langle {q-y}|{\hat A}|{q+y}\rangle$} =
  \int\frac{{\rm d}^{D}{k}\,}{(2\pi\hbar)^D}
  P_{\hat A}(q,k) \,
  \exp\left(-2\frac{i}{\hbar} k\cdot y - \frac{ie}{\hbar}
            \int_{q-y}^{q+y} {\cal A}(x)\cdot{\rm d}{x}\right).
\end{equation}
The matrix representation and the phase-space representation of the
operator $\hat A$ are therefore equivalent.

The Wigner transform (\ref{Wig31}) conserves the linear structure of the
underlying operator algebra
\begin{equation} \label{Wig44}
  P_{\hat A+\hat B}(q,p) = P_{\hat A}(q,p) + P_{\hat B}(q,p),
  \qquad
  P_{\alpha\hat A} = \alpha P_{\hat A}.
\end{equation}
If the operators carry internal degrees of freedom, the Wigner function will
acquire a matrix structure, and the second equation in (\ref{Wig44}) holds
for constant matrices $\alpha$. Because of noncommutativity, multiplication
of operators translates into a more complex phase-space operation considered
below.

The Wigner transforms of the position operator $\hat q$ and the
momentum operator $\hat p$ can be determined directly from
their matrix elements,
\begin{equation} \label{Wig12}
  \mbox{$\langle {q}|{\hat q}|{q'}\rangle$} = \delta(q'-q), \qquad
  \mbox{$\langle {q}|{\hat p}|{q'}\rangle$}
= -i\hbar\,\delta(q'-q) \,\frac{\partial}{\partial (q'-q)}.
\end{equation}
For the position operator, insertion into (\ref{Wig31}) yields
\begin{equation}
  P_{\hat q}(q,p) = q.
\end{equation}
This relation may be generalized to analytic functions  $W(\hat q)$:
$P_{W(\hat q)}(q,p) = W(q)$ by a Taylor expansion. In the case of the
momentum operator, the derivative in (\ref{Wig12}) gives an additional term
from its action on the line integral, which yields
\begin{equation}
  P_{\hat p}(q,p) = p + e{\cal A}(q).
\end{equation}
Therefore, the kinetic momentum $\hat\pi = \hat p-e{\cal A}(\hat q)$ has the
Wigner transform
\begin{equation}
  P_{\hat\pi}(q,p) = p.
\end{equation}
This allows us to interpret the coordinate $p$ as a true (i.e., kinetic)
momentum.

\subsection{Noncommutative multiplication in phase space}
The noncommutative multiplication of the operator algebra translates into
a `noncommutative multiplication' of their Wigner transforms. Consider
\begin{eqnarray}
  P_{\hat A\hat B}(q,p) &=&
    \int {\rm d}^{D}y\,
    \mbox{$\langle {q-\mbox{$\scriptstyle 1\over 2$} y}|
  {\hat A\hat B}|{q+\mbox{$\scriptstyle 1\over 2$} y}\rangle$} \nonumber\\
    && \quad\times
    \exp\left(\frac{i}{\hbar} p\cdot y + \frac{ie}{\hbar}
              \int_{q-y/2}^{q+y/2} {\cal A}(x)\cdot{\rm d}{x}\right).
\end{eqnarray}
Inserting a complete set of intermediate states and using (\ref{invWig31})
gives us---after some shifts of integration variables---the multiplication
formula in its integral form:
\begin{eqnarray} \label{Wig41} &&
  P_{\hat A\hat B}(q,p) =
  (\pi\hbar)^{-2D}
  \int {\rm d}^{D}y\,{\rm d}^{D}y'\,{\rm d}^{D}k\,{\rm d}^{D}k'\,
  \nonumber\\ && \qquad\times
  P_{\hat A}(q+y,p+k') P_{\hat B}(q+y',p+k)
  \nonumber\\ && \qquad\times
  \exp\left\{\frac{2i}{\hbar} (yk - y'k') \right\}
  \nonumber\\ && \qquad\times
  \exp\left\{\frac{ie}{\hbar}
    \left(\,
      \int_{q+y-y'}^{q-y+y'} +
      \int_{q+y+y'}^{q+y-y'} +
      \int_{q-y+y'}^{q+y+y'}\,
    \right)
    {\cal A}(x)\cdot{\rm d}{x}
  \right\}.
\end{eqnarray}
With ${\cal A}(x)$ put to zero one gets the multiplication
formula given by \cite{Hil}.

Let us now consider the usual case $D=3$ with ${\cal A}(x)$ being a
three-dimensional vector potential. The closed-line integral in (\ref{Wig41}
)is related to the surface integral over the magnetic field ${\cal B}(q)$:
\begin{equation}
  \int_{\partial\Delta} {\cal A}(x)\cdot{\rm d}{x} =
  \int_\Delta {\rm d}^{2}x\, n\cdot {\cal B}(x),
  \qquad {\cal B}(q)=\mathop{\rm curl} A(q),
\end{equation}
where $\Delta$ represents the integration region (here a triangle)
and $\partial\Delta$ its border, $n$ is the normal to the triangle plane,
${\rm d}{x}$ the line element of the border, and ${\rm d}^{2}x\,$ the area
element
of the triangle. We parametrize the surface of the triangle according to
\begin{equation}
   x = q+\lambda_1 y +\lambda_2 y', \quad
  -1 \le \lambda_1, \lambda_2 \le 1, \quad
  \lambda_1+\lambda_2 \ge 0.
\end{equation}
The surface integral in question may be rewritten:
\begin{equation}
  \int_\Delta{\rm d}^{2}x\, n\cdot\mathop{\rm curl} {\cal A}(x) =
  \int_{-1}^1 {\rm d}{\lambda_1}
  \int_{-\lambda_1}^{1} {\rm d}{\lambda_2}
  (y\times y')\cdot {\cal B}(q+\lambda_1 y + \lambda_2 y').
\end{equation}

Inserting the above formulae into (\ref{Wig41}) we get
\begin{eqnarray} &&
  P_{\hat A\hat B}(q,p) =
  (\pi\hbar)^{-6}
  \int{\rm d}^{3}y\,{\rm d}^{3}y'\,{\rm d}^{3}k\,{\rm d}^{3}k'\,
  P_{\hat A}(q+y,p+k')
  \nonumber\\ && \quad\times
  \exp\left\{\frac{2i}{\hbar} (yk - y'k')
    +\frac{ie}{\hbar} \, (y\times y')\cdot
     \int_\Delta {\rm d}{\lambda_1}{\rm d}{\lambda_2}
     {\cal B}(q+\lambda_1 y + \lambda_2 y')
  \right\}
  \nonumber\\ && \quad\times
  \exp\left(y'\cdot\frac{\partial}{\partial\tilde q}
  + k\cdot\frac{\partial}{\partial\tilde p}\right)
  \nonumber\\ && \quad\times
  P_{\hat B}(\tilde q,\tilde p)\bigg|_{\tilde q=q,\tilde p=p}
  \quad,
\end{eqnarray}
where we made use of the `translation' formula
\begin{equation}
  P_{\hat B}(q+y',p+k')
  = \left.
    \exp\left( y'\cdot\frac{\partial}{\partial\tilde q}
   + k\cdot\frac{\partial}{\partial\tilde p} \right)
    P_{\hat B}(\tilde q,\tilde p) \right|_{\tilde q=q,\tilde p=p}.
\end{equation}

Now the integration over $k$ can be performed, yielding a
(formal) delta function
\begin{equation}
  \delta\left( y + \frac{\hbar}{2i}\frac{\partial}{\partial\tilde p}
\right).
\end{equation}
Subsequent integration over $dy$ results in the expression:
\begin{eqnarray}
  P_{\hat A\hat B}(q,p) &=&
  \pi^{3}
  \int{\rm d}^{3}y'\,{\rm d}^{3}k'\,
  P_{\hat A}\left(q-\frac{\hbar}{2i}
\frac{\partial}{\partial\tilde p},p+k'\right)
  \nonumber\\ && \quad\times
  \exp\left\{ie \epsilon_{ijk}
    \frac{1}{2i} \frac{\partial}{\partial\tilde p_i} y'_j
    \int_\Delta {\rm d}{\lambda_1}{\rm d}{\lambda_2}
    {\cal B}\left(q-\lambda_1 \frac{\hbar}{2i}
\frac{\partial}{\partial\tilde p}
    + \lambda_2 y'\right)
  \right\}
  \nonumber\\ && \quad\times\left.
  \exp\left\{-\frac{2i}{\hbar}\,
             y'\cdot\left(k' - \frac{\hbar}{2i}
\frac{\partial}{\partial\tilde q}\right)
  \right\}
  P_{\hat B}(\tilde q,\tilde p)\right|_{\tilde q=q,\tilde p=p}.
\end{eqnarray}
We can now get rid of the $y'$ in the first exponential by substituting
it by
\begin{equation}
  y' \to -\frac{\hbar}{2i} \frac{\partial}{\partial k'},
\end{equation}
where the derivative acts on the last exponential. This leads to the
expression
\begin{eqnarray} &&
  P_{\hat A\hat B}(q,p) =
  \int{\rm d}^{3}k'\,
  P_{\hat A}\left(q-\frac{\hbar}{2i}
\frac{\partial}{\partial\tilde p},p+k'\right)
  \nonumber\\ && \quad\times
  \exp\left\{-\frac{e\hbar}{4i} \epsilon_{ijk}
    \frac{\partial}{\partial p_i}\frac{\partial}{\partial k'_j}
    \int_\Delta {\rm d}{\lambda_1}{\rm d}{\lambda_2}
    {\cal B}_k\left(q-\lambda_1 \frac{\hbar}{2i}
\frac{\partial}{\partial\tilde p}
-\lambda_2 \frac{\hbar}{2i} \frac{\partial}{\partial k'}\right)
  \right\}
  \nonumber\\ && \quad\times \left.
  \delta\left(k' - \frac{1}{2i}
\frac{\partial}{\partial\tilde q}\right) \,
  P_{\hat B}(\tilde q,\tilde p)\right|_{\tilde q=q,\tilde p=p}.
\end{eqnarray}
After partial integration, this operator acts
on $P_{\hat A}$ and can be replaced by $-\hbox{$\partial$
  \hbox to0pt{\kern-1em \raise1.5ex\hbox{$\leftarrow$}\hss}}/\partial p$,
where
the arrow indicates that the momentum derivative acts on the function to its
left yielding the final expression
\begin{eqnarray} &&
  P_{\hat A\hat B}(q,p) =
  P_{\hat A}\left(q-\frac{\hbar}{2i}
\frac{\partial}{\partial\tilde p},
p+\frac{\hbar}{2i}\frac{\partial}{\partial\tilde q}\right)
  \nonumber\\ && \quad\times
  \exp\left\{-\frac{e\hbar}{4i} \epsilon_{ijk}
    \frac{\hbox{$\partial$
  \hbox to0pt{\kern-1em \raise1.5ex\hbox{$\rightarrow$}\hss}}}{\partial
                                                                 \tilde p_i}
\frac{\hbox{$\partial$
  \hbox to0pt{\kern-1em \raise1.5ex\hbox{$\leftarrow$}\hss}}}{\partial p_j}
    \int_{-1}^{1} {\rm d}{\lambda_1}
    \int_{-\lambda_1}^{1} \!\!\!{\rm d}{\lambda_2}
    {\cal B}_k\left(q-\lambda_1 \frac{\hbar}{2i}
\frac{\hbox{$\partial$
  \hbox to0pt{\kern-1em \raise1.5ex\hbox{$\rightarrow$}\hss}}}{\partial\tilde
p}
                     +\lambda_2 \frac{\hbar}{2i}
\frac{\hbox{$\partial$
  \hbox to0pt{\kern-1em \raise1.5ex\hbox{$\leftarrow$}\hss}}}{\partial
p}\right)
  \right\}
  \nonumber\\ && \quad\times \left.
  P_{\hat B}(\tilde q,\tilde p)\right|_{\tilde q=q,\tilde p=p}.
\label{mull}
\end{eqnarray}
In this way the left multiplication with an operator $A$ was translated into a
phase-space operation. For the right multiplication we get
\begin{eqnarray} &&
  P_{\hat B\hat A}(q,p) =
  P_{\hat A}\left(q+\frac{\hbar}{2i}
\frac{\partial}{\partial\tilde p},
 p-\frac{\hbar}{2i}\frac{\partial}{\partial\tilde q}\right)
  \nonumber\\ && \quad\times
  \exp\left\{\frac{e\hbar}{4i} \epsilon_{ijk}
    \frac{\hbox{$\partial$
  \hbox to0pt{\kern-1em \raise1.5ex\hbox{$\rightarrow$}\hss}}}{\partial
                                                                \tilde p_i}
 \frac{\hbox{$\partial$
  \hbox to0pt{\kern-1em \raise1.5ex\hbox{$\leftarrow$}\hss}}}{\partial p_j}
    \int_{-1}^{1} {\rm d}{\lambda_1}
    \int_{-1}^{-\lambda_1} {\rm d}{\lambda_2}
    {\cal B}_k\left(q-\lambda_1 \frac{\hbar}{2i}
\frac{\hbox{$\partial$
  \hbox to0pt{\kern-1em \raise1.5ex\hbox{$\rightarrow$}\hss}}}{\partial\tilde
p}
                     +\lambda_2 \frac{\hbar}{2i}
\frac{\hbox{$\partial$
  \hbox to0pt{\kern-1em \raise1.5ex\hbox{$\leftarrow$}\hss}}}{\partial
p}\right)
  \right\}
  \nonumber\\ && \quad\times \left.
  P_{\hat B}(\tilde q,\tilde p)\right|_{\tilde q=q,\tilde p=p}.
\label{mulr}
\end{eqnarray}

\subsection{Kinetic energy and the flow term}
In the following we will derive equations of motion for the Wigner transform
of the density matrix $\hat\rho$. The time evolution of the latter is
governed by the Heisenberg equation of motion
\begin{equation} \label{heqm}
  \frac{\partial \hat\rho}{\partial t} = -\frac{i}{\hbar}[\hat H,\hat\rho].
\end{equation}
Therefore we have to find the phase-space analog of $[\hat H,\hat\rho]$.
We do not assume any special properties of $\hat\rho$, and
the following would apply to a general Heisenberg operator that has no
explicit time dependence. (Allowing for an explict time dependence
does not pose any problem.)

We begin with the kinetic momentum $\hat \pi=\hat p - eA(\hat q)$.  Its
Wigner transform is $P_{\hat\pi}(q,p) = p$. In this case only first-order
derivatives $\hbox{$\partial$
  \hbox to0pt{\kern-1em \raise1.5ex\hbox{$\leftarrow$}\hss}}
/\partial p$ in (\ref{mull}) and (\ref{mulr})
lead to a nonzero result. Therefore only the first two terms
in the Taylor expansion of the exponential are to be left
\begin{eqnarray} &&
  \exp\left\{\frac{e\hbar}{4i} \epsilon_{ijk}
    \frac{\hbox{$\partial$
  \hbox to0pt{\kern-1em \raise1.5ex\hbox{$\rightarrow$}\hss}}}{\partial
                                                                 \tilde p_i}
\frac{\hbox{$\partial$
  \hbox to0pt{\kern-1em \raise1.5ex\hbox{$\leftarrow$}\hss}}}{\partial p_j}
    \int_{-1}^{1} {\rm d}{\lambda_1}
    \int_{-\lambda_1}^{1} {\rm d}{\lambda_2}
    {\cal B}_k\left(q+\lambda_1 \frac{\hbar}{2i}
\frac{\hbox{$\partial$
  \hbox to0pt{\kern-1em \raise1.5ex\hbox{$\rightarrow$}\hss}}}{\partial\tilde
p}
                     -\lambda_2 \frac{\hbar}{2i}
\frac{\hbox{$\partial$
  \hbox to0pt{\kern-1em \raise1.5ex\hbox{$\leftarrow$}\hss}}}{\partial
p}\right)
  \right\} \nonumber\\ &\longrightarrow&
  1 + \frac{e\hbar}{4i} \epsilon_{ijk}
  \frac{\hbox{$\partial$
  \hbox to0pt{\kern-1em \raise1.5ex\hbox{$\rightarrow$}\hss}}}{\partial
                                                                 \tilde p_i}
\frac{\hbox{$\partial$
  \hbox to0pt{\kern-1em \raise1.5ex\hbox{$\leftarrow$}\hss}}}{\partial p_j}
      \int_{-1}^{1} {\rm d}{\lambda_1}
      \int_{-\lambda_1}^{1} {\rm d}{\lambda_2}
      {\cal B}_k\left(q+\lambda_1 \frac{\hbar}{2i}
\frac{\hbox{$\partial$
  \hbox to0pt{\kern-1em \raise1.5ex\hbox{$\rightarrow$}\hss}}}{\partial
                                                               \tilde
p}\right)
\end{eqnarray}
are to be left. Employing this we find
\begin{eqnarray}
  P_{\hat\pi\hat\rho}(q,p) &=& \left\{
  p + \frac{\hbar}{2i} \frac{\partial}{\partial q}
  - \frac{ie\hbar}{4}
    \int_{-1}^1 {\rm d}{\lambda_1} \int_{-\lambda_1}^1 {\rm d}{\lambda_2}
    {\cal B}\left(q-\frac{\lambda_1}{2i}
\frac{\partial}{\partial p}\right)
    \times\frac{\partial}{\partial p}
  \right\}
  \nonumber\\ &&  P_{\hat\rho}(q,p),
  \\
  P_{\hat\rho\hat\pi}(q,p) &=& \left\{
  p - \frac{\hbar}{2i} \frac{\partial}{\partial q}
  + \frac{ie\hbar}{4}
    \int_{-1}^1 {\rm d}{\lambda_1} \int_{-1}^{-\lambda_1} {\rm d}{\lambda_2}
    {\cal B}\left(q-\frac{\hbar\lambda_1}{2i}
\frac{\partial}{\partial p}\right)
    \times\frac{\partial}{\partial p}
  \right\}
  \nonumber\\ &&  P_{\hat\rho}(q,p).
\end{eqnarray}
{}From this one easily obtains the following formulae for the commutator and
anticommutator of $\hat \pi$ and $\hat \rho$:
\begin{eqnarray}
  P_{[\hat\pi,\hat\rho]}(q,p) &=&
  \left\{-i\hbar\frac{\partial}{\partial q}
         -\frac{ie\hbar}{2} \int_{-1}^{1} {\rm d}{\lambda}
          {\cal B}\left(q-\frac{\hbar\lambda}{2i}
\frac{\partial}{\partial p}\right)
          \times \frac{\partial}{\partial p}
  \right\}
  P_{\hat\rho}(q,p), \\
  P_{\{\hat\pi,\hat\rho\}}(q,p) &=&
  \left\{2p
         -\frac{ie\hbar}{2} \int_{-1}^{1} {\rm d}{\lambda} \lambda
          {\cal B}\left(q-\frac{\hbar\lambda}{2i}
\frac{\partial}{\partial p}\right)
          \times \frac{\partial}{\partial p}
  \right\}
  P_{\hat\rho}(q,p).
\end{eqnarray}
It is convenient to define two phase-space operators
$\hat D$ and $\hat P$ :
\begin{eqnarray}
  \hat D &=&
  \left\{\frac{\partial}{\partial q}
         +e \int_{-1/2}^{1/2} {\rm d}{\lambda}
           {\cal B}\left(q+i\hbar\lambda
\frac{\partial}{\partial p}\right)
           \times \frac{\partial}{\partial p}
  \right\} \nonumber\\ &\approx&
  \frac{\partial}{\partial q}
  + e {\cal B}(q)\times
 \frac{\partial}{\partial p}, \label{Wig42}\\
  \hat P &=&
  \left\{p
        -ie\hbar \int_{-1/2}^{1/2} {\rm d}{\lambda} \lambda
         {\cal B}\left(q+i\hbar\lambda
\frac{\partial}{\partial p}\right)
         \times \frac{\partial}{\partial p}
  \right\} \approx p, \label{Wig42a}
\end{eqnarray}
where the expansion to lowest order in $\hbar$ is given. These operators
act on the Wigner transforms in a nonlocal manner. They are generalizations of
the $q$-derivative and the $p$-multiplication in phase-space and their use
simplifies our notation:
\begin{equation} \label{Wig43}
  P_{[\hat\pi,\hat\rho]}(q,p) = -i\hbar\hat D\, P_{\hat\rho}(q,p),
  \qquad
  P_{\{\hat\pi,\hat\rho\}}(q,p) = 2\hat P\, P_{\hat\rho}(q,p).
\end{equation}

We illustrate his technique using a simple example.
Let us consider the Hamiltonian of a nonrelativistic particle coupled
to an external electromagnetic field
\begin{eqnarray}
  \hat H &=& \hat H_{\rm kin} + e{\cal A}^0(q), \label{Wig53} \\
  \hat H_{\rm kin} &=& \frac{(\hat p-eA(q))^2}{2m}
                   = \frac{\hat\pi^2}{2m}
  \to
  P_{\hat H_{\rm kin}}(q,p) = \frac{p^2}{2m},
\end{eqnarray}
where ${\cal A}^0(q)$ is the time-like part of the electromagnetic
four-potential.

To study the time evolution of the Wigner function $P_{\hat\rho}(q,p)$, we
must evaluate the commutator and, in the case of a more complex Hamiltonian
with matrix structure, the anticommutator. Using the relations found above
this is straightforward. We rewrite both as
\begin{eqnarray}
  [\hat\pi^2,\hat\rho] &=& \{\hat\pi,[\hat\pi,\hat\rho]\}, \\
  \{\hat\pi^2,\hat\rho\} &=&
  \mbox{$\scriptstyle 1\over 2$} \{\pi,\{\pi,\rho\}\} +
  \mbox{$\scriptstyle 1\over 2$} [\pi,[\pi,\rho]].
\end{eqnarray}
The double (anti)commutators correspond to a repeated application of the
operators (\ref{Wig42}) and (\ref{Wig42a}):
\begin{eqnarray}
  P_{[\hat\pi^2/2m,\hat\rho]}(q,p) &=&
  -i\hbar\frac{\hat P\cdot\hat D}{m}\, P_{\hat\rho}(q,p), \\
  P_{\{\hat\pi^2/2m,\hat\rho\}}(q,p) &=&
  2\left(\hat P^2 - \frac{\hbar^2}{4}\hat D^2\right) P_{\hat\rho}(q,p).
\end{eqnarray}
In the expression for the commutator we note the classical Liouville flow
term with the minimal coupling to a magnetic field,
\begin{equation}
  \frac{\hat P\cdot\hat D}{m}
  \approx
  \frac{p}{m} \cdot \left(
\frac{\partial}{\partial q} +
e B(q)\times\frac{\partial}{\partial p} \right),
\end{equation}
where only the lowest order term in $\hbar$ was left.

The commutator involving a scalar potential $U(q)$ can also be rewritten as a
differential operator in phase space
\begin{equation} \label{Wig47}
  P_{[\hat U,\hat\rho]}(q,p)
  = i\int_{-1/2}^{1/2} \frac{\partial U}{\partial q}
    \left(q+i\hbar\lambda
\frac{\partial}{\partial p}\right) P_{\hat\rho}(q,p)
  \approx i\hbar
\frac{\partial U}{\partial q} P_{\hat\rho}(q,p) + O(\hbar^2).
\end{equation}
The correspondig expression for the anticommutator is
\begin{eqnarray} \label{Wig49}
  P_{\{\hat U,\hat\rho\}}(q,p)
  &=& U\left(q-\frac{\hbar}{2i}
\frac{\partial}{\partial p}\right) P_{\hat\rho}(q,p)
     +U\left(q+\frac{\hbar}{2i}
\frac{\partial}{\partial p}\right) P_{\hat\rho}(q,p) \nonumber\\
  &\approx& U(q) P_{\hat\rho}(q,p) + O(\hbar^2).
\end{eqnarray}

The time evolution of a Wigner function is not only determined by the
change of the underlying operator, which is usually given by the Heisenberg
equation of motion, but also by the change in the line integral due to a
time-dependent vector potential. The additional term can be rewritten in a
manner similar to that used above:
\begin{equation} \label{Wig46}
  \frac{\partial}{\partial t}P_{\hat\rho}(q,p) =
  P_{\partial_t\hat\rho}
  +e \int_{-1/2}^{1/2}
   \partial_t{\cal A}\left(q+i\hbar\lambda
\frac{\partial}{\partial p}\right)
   \cdot\frac{\partial}{\partial p} \,
   P_{\hat\rho}.
\end{equation}
Inserting the Heisenberg equation of motion (\ref{heqm})
with the Hamilton operator of Eq.~(\ref{Wig53}), we obtain
\begin{equation}
  D_t P_{\hat\rho}(q,p) =
  -\frac{\hat P \cdot \hat D}{m} P_{\hat\rho}(q,p)
\end{equation}
with the generalized time derivative
\begin{equation} \label{Wig54}
  D_t = \frac{\partial}{\partial t}
      + \int_{-1/2}^{1/2} e{\cal E}\left(q+i\hbar\lambda
\frac{\partial}{\partial p}\right)
        \cdot\frac{\partial}{\partial p}
  \approx \frac{\partial}{\partial t}
 + e{\cal E}(q)\cdot\frac{\partial}{\partial p} + \cdots
\end{equation}
The derivatives of ${\cal A}$ resulting from (\ref{Wig47}) and (\ref{Wig46})
were combined to yield an electric field
\begin{equation}
  {\cal E}(q) = -\frac{\partial {\cal A}}{\partial t}
- \frac{\partial {\cal A}^0}{\partial q}.
\end{equation}
To lowest order in $\hbar$ this is the nonrelativistic Liouville equation
of a charged particle in an external electromagnetic field.

\subsection{Calculation of observables}
The Wigner function is normalized to give
\begin{equation}
  \int{\rm d}^{D}q\,\frac{{\rm d}^{D}p\,}{(2\pi\hbar)^D}\,
  P_{\hat A}(q,p) = \int{\rm d}^{D}p\,
  \mbox{$\langle {q}|{\hat A}|{q}\rangle$}
  \equiv \mathop{\rm Tr}\hat A.
\end{equation}
In the case of a density matrix it is just one. Observables are computed by
taking the trace over the product of density matrix and observable operator:
Using the multiplication formula (\ref{Wig41}) one can prove that
\begin{equation} \label{obs}
  \langle \hat O \rangle \equiv \mathop{\rm Tr}\left(\hat O\,\hat\rho\right) =
  \int{\rm d}^{D}q\, \frac{{\rm d}^{D}p\,}{(2\pi\hbar)^D}\,
  \mathop{\rm Tr}\left( P_{\hat O}(q,p) P_{\hat\rho}(q,p) \right),
\end{equation}
where $\mathop{\rm Tr}$ is the trace over the matrix structure of $P(q,p)$ (if
necessary). This formula allows us to compute any observable from the
phase-space representation of the density matrix; it further suggests the
identification of the phase-space function of an observable operator and the
classical observable function.

\section{The Wigner function of the Klein--Gordon field}  \label{secWigner}
We will now treat the equal-time Wigner function of a Klein--Gordon field in
an external electromagnetic field. Instead of the matrix elements of a
density matrix, the object to consider will be the symmetrized two-point
correlation function. Its properties are very similar to a
quantum-statistical density matrix in that it describes a many-particle
system. All single-particle observables can be computed from the two-point
function, and thus from its Wigner transform.

The time evolution of the two-point function is derived from the
Klein--Gordon equation for the Heisenberg field operators. Similarly to the
nonrelativistic problem treated above the Klein-Gordon equation  can be
translated into a differential equation in phase-space with its leading
contribution being just the classical relativistic flow equation. Two
complications arise:  The Klein--Gordon equation contains a second-order
derivative in time. Its translation into phase-space would also contain a
second-order derivative which does not lead to a sensible evolution
equation. The reason for this is that the Klein--Gordon field has an
internal charge degree of freedom which can be made explicit in the
two-component formalism of Feshbach and Villars \cite{FV}. In this
representation, the time evolution is first-order in time and governed by a
nonhermitian Hamilton operator with a $2\times2$-matrix structure. As a
second complication in a fully interacting case (scalar electrodynamics) the
equation of motion of the two-point function involves higher correlation
functions. Here we consider the simpler problem of the interaction with a
classical albeit dynamical electromagnetic field in which case the equation
of motion for the correlation function is closed and the electromagnetic
field is governed by the Maxwell equations.

\subsection{Definition}
The Klein--Gordon field which obeys the wave equation ($\hbar=c=1$)
\begin{equation} \label{Wig56}
  \left((\partial_\mu - ie A_\mu)
        (\partial^\mu - ie A^\mu) + m^2\right) \phi(x) = 0,
\end{equation}
can be expressed by a two-component wave function
\begin{equation}
  \Phi = \left(\begin{array}{c}\psi \\ \chi\end{array}\right)
\end{equation}
transforming regularly under U(1)-gauge transformations where
\begin{equation}
  \psi = \frac{1}{2}\left( \phi + \frac{i}{m}
\frac{\partial\phi}{\partial t}
                      - \frac{e{\cal A}^0}{m} \phi \right),
  \qquad
  \chi = \frac{1}{2}\left( \phi - \frac{i}{m}
\frac{\partial\phi}{\partial t}
                      + \frac{e{\cal A}^0}{m} \phi \right).
\end{equation}
The equation of motion for the two-component field operator is:
\begin{equation} \label{PVeq}
  i\frac{\partial}{\partial t}\Phi = \left[
  \frac{1}{2m} \left(-i\frac{\partial}{\partial q} - e{\cal A}\right)^2
    \left(\begin{array}{cc} 1 & 1 \\ -1 & -1\end{array}\right)
  + m \left(\begin{array}{cc} 1 & 0 \\ 0 & -1\end{array}\right)
  + e {\cal A}^0 {\newbox\einsbox \setbox\einsbox\vbox{1}
  1\kern-1.7pt\vrule width0.5pt height\ht\einsbox depth0.0pt\relax}
\right]\Phi.
\end{equation}
The advantage of this equation over (\ref{Wig56}) is that the time
derivative appears in the first-order giving rise to a
Schr\"odinger type evolution equation.
The right-hand side can then be interpreted as a Hamiltonian
operator acting on $\Phi$ and the formalism of the previous section can be
utilized to derive a phase-space equation. The fact that the resulting
Hamiltonian operator is not hermitian does not cause any severe problems,
since the nonhermiticity is confined to the matrix structure.

In the following, we consider the time evolution of the symmetrized
correlation function
\begin{equation}
  C_{\alpha\beta}^+(q,q';t)
  = \mbox{$\langle {\Omega}|
{\{\Phi_\alpha(q,t),\Phi_\beta^+(q',t)\}}|{\Omega}\rangle$}
\end{equation}
which is a $2\times2$ matrix. The matrix element is taken
in the Hilbert space of quantum field theory with respect to some state
$\Omega$. Later on this state will be identified with the vacuum. We
may consider $C^+(q,q')$ to be the matrix element
of some density matrix $\hat\rho$ in the abstract Hilbert space
spanned by position eigenvectors $\mbox{$\langle q|$}$,
\begin{equation}
  \mbox{$\langle {q}|{\hat\rho}|{q'}\rangle$} = C^+(q,q').
\end{equation}
By Eq. (\ref{PVeq}), the density matrix $\hat\rho$ evolves under the
Heisenberg equation of motion
\begin{equation} \label{Wig55}
  i \frac{\partial\hat\rho}{\partial t} = \hat H\hat\rho - \hat\rho\hat H^+
\end{equation}
with the two-component nonhermitian Hamiltonian
\begin{equation} \label{Wig45}
  \hat H = \frac{(\hat{p} - e{\cal A})^2}{2m}
\left(\begin{array}{cc} 1 & 1 \\ -1 & -1\end{array}\right)
           + m \left(\begin{array}{cc} 1 & 0 \\ 0 & -1\end{array}\right)
                  + e {\cal A}^0 {\newbox\einsbox \setbox\einsbox\vbox{1}
  1\kern-1.7pt\vrule width0.5pt height\ht\einsbox depth0.0pt\relax},
\end{equation}
where $\hat p$ denotes momentum operator $-i\partial/\partial q$.

The energy, momentum, charge, and current of the Klein--Gordon field
are bilinear forms in the field operators that can be
computed from the two-point correlation function or from the Wigner
function. For example, the energy and momentum
of free particles are given by the elements of the energy-momentum tensor,
\begin{eqnarray} \label{Wig8}
  \Theta^{00} &=& \frac{1}{2m}
                  \left(\frac{\partial\phi}{\partial t}\right)^*
                  \left(\frac{\partial\phi}{\partial t}\right) +
                  (\nabla\phi^*)(\nabla\phi) + m^2 \phi^*\phi \\
              &=& \Phi^+ \left[
\left(\begin{array}{cc} 1 & 1 \\ 1 & 1\end{array}\right)
                                \frac{\hat p^2}{2m} +
\left(\begin{array}{cc} 1 & 0 \\ 0 & 1\end{array}\right) m
                                \right] \Phi, \\
\Theta^{0i} &=& \frac{\partial\phi^*}{\partial t} \nabla\phi
                 +\nabla\phi^* \frac{\partial\phi}{\partial t} \\
              &=& \Phi^+ \left[ -m
\left(\begin{array}{cc} 1 & 1 \\ -1 & -1\end{array}\right)
                                   \hat p \right] \Phi.
\end{eqnarray}
The sum of these quantities and the corresponding quantities of the
electromagnetic field are conserved. Charge and current density operators
read
\begin{eqnarray} \label{Wig9}
  \rho_e &=& \frac{ie}{2m} \left(\phi^*
\frac{\partial\phi}{\partial t}
 -\phi\frac{\partial\phi^*}{\partial t}\right)
             - \frac{e^2 A^0}{m} \phi^*\phi
          =  \Phi^+ \left[ e
\left(\begin{array}{cc} 1 & 0 \\ 0 & -1\end{array}\right)
 \right] \Phi, \\
  j      &=& \frac{e}{2mi} \left(\phi^*\nabla\phi -
                                  \phi\nabla\phi^*\right)
             - \frac{e^2 A}{m} \phi^*\phi
          =  \Phi^+ \left[ \frac{e}{m} \hat p
\left(\begin{array}{cc} 1 & 1 \\ 1 & 1\end{array}\right)
                           \right] \Phi.
\end{eqnarray}
The expressions in brackets constitute the observable operators in the sense
of Eq.~(\ref{obs}).

\subsection{The Wigner function and its time evolution}
The Wigner transform associates a matrix-valued phase-space function
$P(q,p)$ with the density matrix $\hat\rho$. Its evolution is governed by
Eq. (\ref{Wig55}) where the nonhermiticity of the Hamiltonian (\ref{Wig45})
has been taken into account. This may be translated into a phase-space
evolution equation. Defining
\begin{equation}
  a = (\sigma_3 + i\sigma_2) =
\left(\begin{array}{cc} 1 & 1 \\ -1 & -1\end{array}\right), \qquad
  b = \sigma_3 = \left(\begin{array}{cc} 1 & 0 \\ 0 & -1\end{array}\right)
\end{equation}
and using (\ref{Wig46}) one obtains:
\begin{eqnarray}
  P_{\hat H\hat\rho - \hat\rho\hat H^+}(q,p) &=&
  \frac{1}{4m} \left(a\cdot P_{[\hat p^2,\hat\rho]}(q,p)
                     +P_{[\hat p^2,\hat\rho]}(q,p)\cdot a^+\right)
  \nonumber\\ &&
  {}+\frac{1}{4m} \left(a\cdot P_{\{\hat p^2,\hat\rho\}}(q,p)
                     -P_{\{\hat p^2,\hat\rho\}}(q,p)\cdot a^+\right)
  \nonumber\\ &&
  {}+m \left(b\cdot P(q,p) - P(q,p)\cdot b^+\right).
\end{eqnarray}
Following the general procedure outlined in the previous section
we get the phase-space equation of motion:
\begin{eqnarray} \label{Wig6}
  i D_t P(q,p) &=&
  {}-i \frac{\hat P}{2m}\cdot\hat D
   \left( a\cdot P(q,p) + P(q,p) \cdot a^+ \right) \nonumber\\ &&
  {}+\frac{1}{m}\left( \hat P^2 - \frac{1}{4} \hat D^2 \right)
   \left( a\cdot P(q,p) - P(q,p) \cdot a^+ \right) \nonumber\\ &&
  {}+ m\left(b\cdot P(q,p) - P(q,p) \cdot b\right).
\end{eqnarray}

It is convenient to expand $P(q,p)$ (which is a $2\times 2$ matrix) over the
Pauli matrices $\sigma_i$ and the unit matrix, assembling the coefficients
into a four-component object $\underline{f}\,$,
\begin{equation} \label{Wig50}
  P(q,p) = f_3(q,p) \,{\newbox\einsbox \setbox\einsbox\vbox{1}
  1\kern-1.7pt\vrule width0.5pt height\ht\einsbox depth0.0pt\relax}
 + \sum_{i=1}^3 f_{3-i}(q,p,t) \sigma_i.
\end{equation}
The equations of motion take the form
\begin{eqnarray} \label{Wig18}
  \hat D_t f_0 &=& -\frac{\hat P\cdot\hat D}{m} (f_2+f_3), \nonumber\\
  \hat D_t f_1 &=&
    -\left(\frac{1}{4m} \hat D^2 - \frac{1}{m} \hat P^2\right)
    (f_2+f_3)
    + 2m f_2, \nonumber\\
  \hat D_t f_2 &=&
    \left(\frac{1}{4m} \hat D^2 - \frac{\hat P^2}{m} \right) f_1
    +\frac{\hat P\cdot\hat D}{m} f_0
    - 2m f_1, \nonumber\\
  \hat D_t f_3 &=&
    -\left(\frac{1}{4m} \hat D^2 - \frac{\hat P^2}{m}\right) f_1
    -\frac{\hat P\cdot\hat D}{m} f_0.
\end{eqnarray}

The components of $\underline{f}\,$ may be interpreted as the phase-space
densities of
simple observables:
\begin{eqnarray}
  \mbox{charge:} &&
  ef_0, \\
  \mbox{energy:} &&
  \frac{p^2}{2m} (f_2+f_3) + m f_3, \\
  \mbox{current:} &&
  \frac{p}{m} e (f_2+f_3), \\
  \mbox{momentum:} &&
  p (f_0 - f_1).
\end{eqnarray}

It may be interesting to note the conservation law that follows
from the preservation of the norm
\begin{equation}
  \frac{\partial}{\partial t}
 \int\frac{{\rm d}^{D}q\,{\rm d}^{D}p\,}{(2\pi\hbar)^3}
  (f_0^2 - f_1^2 - f_2^2 + f_3^2)
  = 0.
\end{equation}
The Feshbach-Villars Hamiltonian (\ref{Wig45}) does not give rise to unitary
evolution but to $\tau$-unitary evolution \cite{Davydov} so that two $f$s
come in with the opposite sign. The conservation laws for energy-momentum
and currents may be easily derived and we do not state them here.

The free vacuum solution can be derived from the explicit expression
for the correlation function $C^+$:
\begin{eqnarray}
  f_0 = 0, &\quad& f_1=0, \nonumber\\
  f_2+f_3 = \frac{m}{E_p}, &\quad&
  f_3-f_2 = \frac{E_p}{m}.
\label{VacSol}
\end{eqnarray}
It is the stationary solution of (\ref{Wig18}).
The vacuum has a constant energy density (which must be subtracted when
computing observables) but no charge.


\subsection{Diagonalization of local oscillations}
For a further discussion of these equations we consider the case of a
electric field slowly varying in space. This will allow us to drop the
higher-order corrections in $\hat P$ and $\hat D$. Nevertheless, the
equation of motion (\ref{Wig18}) contains self-couplings of the components
that make its solutions nonstationary even in the absence of a gradient. The
resulting oscillations are related to the internal charge degree of
freedom. The eigenfrequencies of these oscillations are (in the absence of
an electromagnetic field) just $\pm 2E_p$, showing clearly that they are
related to relativistic effects and therefore to the existence of
antiparticles. Let us expand the Wigner function $P(q,p)$ into eigenmodes
$\tilde P_i$ of these oscillations:
\begin{equation}
  P(q,p) = \sum_{i=1}^4 \tilde f_i \tilde P_i,
\end{equation}
with
\begin{eqnarray} \label{Wig48}
  \tilde P_1 &=& \frac{1}{4} \left(\begin{array}{cc}
                 m/E_p + E_p/m    & m/E_p - E_p/m    \\
                 m/E_p - E_p/m    & m/E_p + E_p/m
                 \end{array}\right), \nonumber\\
  \tilde P_2 &=& \frac{1}{2} \left(\begin{array}{cc}
                 1                & 0 \\
                 0                & 1
                 \end{array}\right), \nonumber\\
  \tilde P_{3/4} &=& \frac{1}{4} \left(\begin{array}{cc}
                     m/E_p - E_p/m         & m/E_p + E_p/m \mp 1/2 \\
                     m/E_p - E_p/m \pm 1/2 & m/E_p - E_p/m
                   \end{array}\right).
\end{eqnarray}
The first two matrices are connected with the eigenfrequency zero while
the last two correspond to $\pm2E_p$. Each of these matrices can be
constructed from plane-wave solutions of the Klein--Gordon equation. Note
that the basis matrices $\tilde P_i$ depend on the energy $E_p$ and are
therefore different in different parts of the phase space. A particle whose
inner degree of freedom is oscillating in an eigenmode of these equations
will be out of phase when it gains or loses momentum.

With the approximation
\begin{equation}
  p^2 + m^2 + \frac{1}{4} \frac{\partial^2}{\partial^2 q}
  \approx p^2 + m^2,
\end{equation}
which is justified in the semiclassical case, the equations of motion take
on the form
\begin{eqnarray}\label{Wig21}
  \left(
\frac{\partial}{\partial t} \tilde f_1 +
e{\cal E}(q)\cdot\frac{\partial}{\partial p}\right)
  &=& -\frac{p}{E_p} \cdot\frac{\partial}{\partial q}\tilde f_2
      + e\frac{{\cal E}(q)\cdot p}{E_p^2} (\tilde f_3 + \tilde f_4), \\
  \left(\frac{\partial}{\partial t}
 \tilde f_2 + e{\cal E}(q)\cdot\frac{\partial}{\partial p}\right)
  &=& -\frac{p}{E_p} \cdot\frac{\partial}{\partial q}
                         \left(\tilde f_1+\tilde f_3+\tilde f_4\right), \\
  \left(
\frac{\partial}{\partial t} \tilde f_3 + e{\cal E}(q)\cdot
\frac{\partial}{\partial p}\right)
  &=& -\frac{1}{2}\frac{p}{E_p} \cdot
\frac{\partial}{\partial q} \tilde f_2 \nonumber\\ &&
      {}+ e\frac{{\cal E}(q)\cdot p}{E_p^2} \frac{\tilde f_1}{2}
      + 2iE_p \tilde f_3, \label{Wig21c}\\
  \left(\frac{\partial}{\partial t} \tilde f_4 +
e{\cal E}(q)\cdot\frac{\partial}{\partial p}\right)
  &=& -\frac{1}{2}\frac{p}{E_p} \cdot
\frac{\partial}{\partial q} \tilde f_2 \nonumber\\ &&
      {}+ e\frac{{\cal E}(q)\cdot p}{E_p^2} \frac{\tilde f_1}{2}
      - 2iE_p \tilde f_4. \label{Wig21d}
\end{eqnarray}
Here $\tilde f_1$, $\tilde f_2$ and the linear combination $\tilde f_3 +
\tilde f_4$
are real, while $\tilde f_3-\tilde f_4$ is purely imaginary.

These variables have the meaning of the following phase-space densities:
\begin{eqnarray}
  \mbox{charge:} && e \tilde f_2, \label{Wig22} \\
  \mbox{energy:} && E_p \tilde f_1 \\
  \mbox{current:} && \frac{pe}{E_p} (\tilde f_1 + \tilde f_3 + \tilde f_4), \\
  \mbox{momentum:} && -p \tilde f_2
                    +ip (\tilde f_3 - \tilde f_4).
\end{eqnarray}
The Wigner transform of a positive- or negative-energy plane-wave solution
with the momentum $p_0$ of the Klein--Gordon equation is
\begin{equation}
  \tilde f_1 = 1 \,\delta(p-p_0), \qquad
  \tilde f_2 = \pm 1 \,\delta(p-p_0), \qquad
  \tilde f_3 = \tilde f_4 = 0.
\end{equation}
Wigner transforms with nonzero $\tilde f_3$ and $\tilde f_4$ involve
interferences between positive- and negative-energy modes
(``Zitterbewegung''). The correlation function of a free vacuum may be
expressed as the superposition of all positive- and negative-energy
contributions (with the same sign, as opposed to the Feynman propagator). The
free vacuum solution is therefore
\begin{equation}
  \tilde f_1 = 1, \quad
  \tilde f_2 = \tilde f_3 = \tilde f_4 = 0.
\end{equation}

In Eq.~(\ref{Wig21})--(\ref{Wig21d}), the classical limit is exhibited in a
more elegant way.  The terms on the right-hand sides of Eqs.~(\ref{Wig21c})
and (\ref{Wig21d}) are comparable only if the variation of $f_2$ is on a
scale of $E_p$, i.e.~the Compton wavelength. In a classical limit, where
this is not the case, the last two equations decouple since $\tilde f_3$ and
$\tilde f_4$ will vary so fast that their contribution to $\tilde f_2$ is
averaged
out. By introducing two new quantities
\begin{eqnarray} \label{Wig32}
  f_+ = \mbox{$\scriptstyle 1\over 2$}(\tilde f_1 + \tilde f_2), &\qquad&
\tilde f_1 = f_+ + f_-,\nonumber\\
  f_- = \mbox{$\scriptstyle 1\over 2$}(\tilde f_1 - \tilde f_2), &\qquad&
\tilde f_2 = f_+ - f_-,
\end{eqnarray}
which obey
\begin{eqnarray}
  \frac{\partial}{\partial t} f_+
  &=& -\frac{p}{E_p} \cdot\frac{\partial}{\partial q}f_+, \\
  \frac{\partial}{\partial t} f_-
  &=& \frac{p}{E_p} \cdot\frac{\partial}{\partial q} f_-,
\end{eqnarray}
one obtains two decoupled Vlasov equations of motion. As can be seen from
(\ref{Wig22}), $f_+$ and $f_-$ carry different charges and thus describe the
phase-space densities of positive and negative particles. In this case
we have a collisionless relativistic gas consisting of positively and
negatively charged particles.

Going back to the equations (\ref{Wig21})--(\ref{Wig21d}) we can give some
meaning to the individual terms. As we have seen $f_1$ and $f_2$ are the
particle and charge number density and therefore represent the `substantial'
part of the Wigner function. Quantities $f_3$ and $f_4$ arise from
interferences between these two parts. On the right-hand side, the local
derivatives involving $\tilde f_1$ and $\tilde f_2$ correspond to flow  while
those involving $\tilde f_3$ and $\tilde f_4$ implement local interference
(together with the neglected second derivative). Most important, the terms
proportional to
\begin{equation}
  e\frac{e{\cal E}(q)\cdot p}{E_p^2}
\end{equation}
stem solely from the momentum dependence of the basis (\ref{Wig48}). They
give rise to pair creation \cite{Schw}. This will be discussed in detail in
a future publication \cite{next}.

\section{Symmetry breaking}                                 \label{secSSB}
As mentioned in section \ref{secIntro}, the phase-space description may
be useful to study the back reaction problems. We would like to
ilustrate this point with a relatively simple example. Namely,
we now show that the formalism developed above
provides a useful tool for the dynamical description of
spontaneous symmetry breaking. The Klein--Gordon equation coupled to a
scalar potential reads
\begin{equation} \label{Wig52}
  (i\partial_t)^2 \phi =
  \left(-i\frac{\partial}{\partial q}\right)^2 \phi + m^2\phi + U(q,t)\phi.
\end{equation}
This leads to the interaction Hamiltonian (in Feshbach-Villars form)
\begin{equation}
  \hat H_{\rm int} = \frac{U(q,t)}{2m} \,
  \left(\begin{array}{cc}  1 & 1 \\ 1 & 1 \end{array}\right).
\end{equation}
Using (\ref{Wig47}) and (\ref{Wig49}) and expanding as in (\ref{Wig50}),
we get
\begin{eqnarray} \label{Wig51}
  \frac{\partial}{\partial t} f_0
&=& -\frac{p}{m} \cdot\frac{\partial}{\partial q} (f_2+f_3)
    + \frac{1}{2m} U_-(q)\,(f_2+f_3) \label{f0eq},\\
  \frac{\partial}{\partial t} f_1 &=&
    -\left(\frac{1}{4m} \frac{\partial^2}{\partial q^2} - \frac{p^2}{m}
\right)
    (f_2+f_3)
    + 2m f_2 \nonumber\\ &&\qquad
    {}+ \frac{1}{2m} U_+(q)\,(f_2+f_3), \\
  \frac{\partial}{\partial t} f_2 &=&
    \left(\frac{1}{4m} \frac{\partial^2}{\partial q^2}
                     - \frac{p^2}{m} \right) f_1
    +\frac{p}{m} \cdot\frac{\partial}{\partial q} f_0
    -2mf_1 \nonumber\\ &&\qquad
    {}- \frac{1}{2m} \left(U_-(q) f_0 + U_+(q) f_1\right),\\
  \frac{\partial}{\partial t} f_3 &=&
    -\left(\frac{1}{4m} \frac{\partial^2}{\partial q^2}
                     - \frac{p^2}{m} \right) f_1
    -\frac{p}{m} \cdot\frac{\partial}{\partial q} f_0 \nonumber\\ && \qquad
    {}+ \frac{1}{2m} \left(U_-(q) f_0 + U_+(q) f_1\right), \label{f3eq}
\end{eqnarray}
with the potentials $U_+$ and $U_-$ given by
\begin{eqnarray}
  U_+(q,t) &=& U\left(q+\frac{i}{2}\frac{\partial}{\partial p},t\right)
            +U\left(q-\frac{i}{2}
\frac{\partial}{\partial p},t\right) \nonumber\\
         &\approx& 2U(q,t) + O(\hbar^2), \\
  U_-(q,t) &=& -i\left[U\left(q+\frac{i}{2}
\frac{\partial}{\partial p},t\right)
                    -U\left(q-\frac{i}{2}
\frac{\partial}{\partial p},t\right)\right]
         \approx 0 + O(\hbar).
\end{eqnarray}

We shall consider the case when the potential $U$ is generated
by the scalar field itself.
To illustrate this point let us consider the charged field $\phi$
described by the Klein--Gordon equation (\ref{Wig52}) with
\begin{equation}
U = - M^2 + \lambda |\phi|^2
\label{SbrkU}
\end{equation}
where the positive constant $\lambda$ defines the strength of
$|\phi|^4$ self-coupling.

It is quite easy to find all time (and coordinate)
independent c-number solutions
to this equation. For $M^2 < m^2$ the only possibility is
\begin{equation}
\phi = 0,
\label{Ssol}
\end{equation}
while for $M^2 > m^2$ there is an additional set of solutions
\begin{equation}
|\phi| = \sqrt{(M^2 - m^2)/\lambda}.
\label{Bsol}
\end{equation}
The solutions (\ref{Bsol}) are known as the symmetry-breaking
solutions.

It is interesting to calculate the Wigner function
corresponding to these classical solutions. The case $\phi = 0$ leads
to the null Wigner function and is of no special interest.
On the other hand the symmetry breaking solution gives rise to
\begin{equation}
f_2(q,p) + f_3(q,p) =
\frac{(M^2-m^2)}{\lambda} \delta^3(p),
\label{SbrkWF}
\end{equation}
while all other components of the Wigner function are equal to zero.
It is easy to show that the above Wigner function satisfies our
basic equations (\ref{f0eq})--(\ref{f3eq}) with the scalar
potential given by
\begin{equation}
U = - M^2 + \lambda \int \frac{d^3p}{(2\pi)^3}
\left( f_2 + f_3 \right).
\label{SbrkU1}
\end{equation}
which follows from Eq. (\ref{SbrkU})
and the definitions of $f_2$ and $f_3$.
Note that this form of potential introduces self-coupling
of a mean-field type.

The appearance of the Dirac delta function reflects the infinite-range
correlation of the classical solution (\ref{Bsol}). The infinite-range
correlation (of quantum fields) indicates symmetry breaking.

The classical considerations presented above do not allow for
the dynamical description of a symmetry-breaking phase transition.
Let us consider a simple---albeit rather abstract---situation in
which one is allowed to switch on the $-M^2$ in (\ref{SbrkU})
at a given time $t=t_0$. (We will assume that it was switched on
instantly but other possibilities are not excluded.)
Initially the quantum
field $\phi$ is in its symmetric vacuum state with mass
equal to $m$. After the external potential $-M^2$ is switched on
this will no longer be a stable state (for suffiuciently large $M$.
Classically one has to provide a small and quite
arbitrary deviation from $\phi =0 $ to initialize the fall
towards the ``true'' vacuum. There is also no way to damp the
resulting oscillations.
Therefore it is not possible to predict the average time
necessary to complete the phase transition from purely classical
considerations. On the other hand quantum theory provides the
fluctuations necessary to drive the transition.
Indeed, equations (\ref{f0eq})--(\ref{f3eq}) contain the necessary
quantum fluctuations and it is possible to study
the dynamics of symmetry breaking using them.

For simplicity we concentrate on the spatially uniform and
isotropic solutions. In that case functions $f$ depends only on
$|p|$; they do depend neither on $q$ nor on the direction
of the momentum $p$. The potential $U$ is also uniform and our
evolution equations (\ref{f0eq})--(\ref{f3eq}) reduce to:
\begin{eqnarray}
  \frac{\partial}{\partial t} f_0 &=& 0 \label{f0uni},\\
  \frac{\partial}{\partial t} f_1 &=&
   \frac{p^2}{m} (f_2+f_3)
   + 2m f_2 + \frac{U(t)}{m}\,(f_2+f_3), \label{f1uni} \\
  \frac{\partial}{\partial t} f_2 &=&
   - \frac{p^2}{m} f_1 -2mf_1
   - \frac{U(t)}{m} f_1,\\
  \frac{\partial}{\partial t} f_3 &=&
   \frac{p^2}{m} f_1
   + \frac{U(t)}{m} f_1. \label{f3uni}
\end{eqnarray}
Note that $f_0$ decouples completely from the rest of the system
and does not undergo any evolution.

The time dependent scalar potential is chosen to be:
\begin{equation}
U(t) = - M^2 \theta(t-t_0) + \lambda \int \frac{d^3p}{(2\pi)^3}
\left( f_2 + f_3 \right) + C.
\label{Self}
\end{equation}
The first term is some external potential that is switched on
(rapidly) at $t=t_0$. If $M^2>m^2$ it will change the sign
of a ``mass squared term'' inducing the transition towards
new equilibrium.
The second term describes a mean field self-coupling of $|\phi^4|$
type. It is well known that for $\lambda >0$ this type of interaction
will tend to stabilise the system.
In addition we have a constant $C$; its value will be specified
in a moment.

As an initial condition we assume the free vacuum state with distribution
functions given by Eq. (\ref{VacSol}) i.e.,
\begin{eqnarray}
  f_0^{t=-\infty} = 0, &\quad& f_1^{t=-\infty}=0, \nonumber\\
  f_2^{t=-\infty}+f_3^{t=-\infty} = \frac{m}{E_p}, &\quad&
  f_3^{t=-\infty}-f_2^{t=-\infty} = \frac{E_p}{m},
\label{VacPast}
\end{eqnarray}
were superscript $t=-\infty$ tells us that we assumed this form
of Wigner function in a distant past.
Note that due to the self-coupling term this state is no longer stable
(in general) and will undergo some evolution even for $t<t_0$ i.e.,
in absence of the external potential $-M^2 \theta(t-t_0)$.
This means that the self-interaction tries to rearrange the free
vacuum which was imposed as an initial state.
It is possible to prevent this rearrangement if we choose
constant $C$ to be:
\begin{equation}
C = - \lambda \int \frac{d^3p}{(2\pi)^3} \frac{m}{E_p}.
\label{RenConst}
\end{equation}
This constant cancels exactly the self-interaction term and the initial
free vacuum becomes a stationary solution to (\ref{f0uni}-\ref{f3uni})
and stays unchanged as long as $t<t_0$.
Note that $C$ is defined by the initial vacuum and stays fixed even
when we switch on an external potential.

The only problem is that the integrals defining $C$ and self-interaction
diverge and one has to introduce a finite cutoff $\Lambda$ to make
the subtraction meanigful. We work with an isotropic system so
the angular integration is not a problem:
\begin{equation}
\int d^3p \,\to\, 4\pi\int_0^\infty d|p|\, p^2 \to
4\pi\int_0^\Lambda d|p|\, p^2.
\end{equation}
The last substitution is just the desired regularization.
The presence of divergences is not a strange here. In fact the
constant $C$ may be traced back to the ``mass renormalization''
counterterm in the Lagrangian.
(There may be some confusion related to the fact that ``mass''
appears not only in $U$ but also in other parts of our expressions.
This is due to Feshbach-Villars transformation. The mass term
plays an important role in this transformation but we are free to use
only a part of it -- the rest may be attached to a scalar potential.
Note, however, that our wave functions and subsequently the Wigner
function are normalized with respect to the mass $m$ i.e., the
one that participates in the Feshbach-Villars transformation.
This must be kept in mind when interpreting the results.)

We solved evolution equations (\ref{f1uni})--(\ref{f3uni}) numerically
for a finite volume of phase space $0\leq |p| <\Lambda$. This automatically
introduces the cutoff. We have chosen it to be rather
large so that we did not observe its effects.

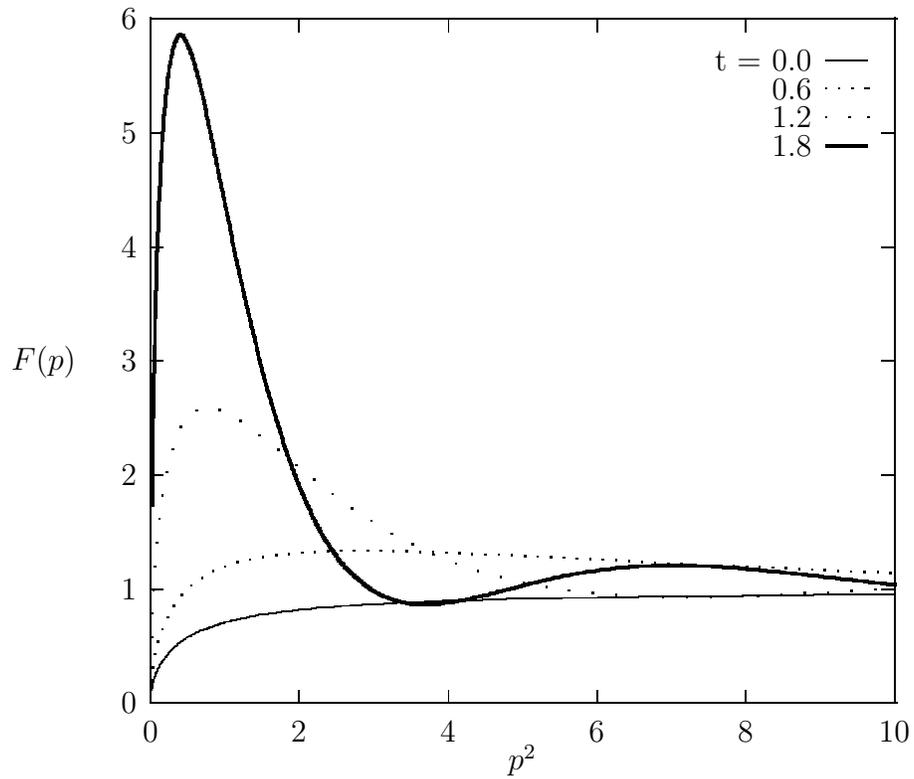
\begin{figure}[th]
\center{
\setlength{\unitlength}{0.24pt}
\ifx\plotpoint\undefined\newsavebox{\plotpoint}\fi
\sbox{\plotpoint}{\rule[-0.175pt]{0.350pt}{0.350pt}}%
\begin{picture}(1500,1350)(0,0)
\normalsize\rm
\put(264,158){\rule[-0.175pt]{282.335pt}{0.350pt}}
\put(264,158){\rule[-0.175pt]{0.350pt}{259.931pt}}
\put(264,158){\rule[-0.175pt]{4.818pt}{0.350pt}}
\put(242,158){\makebox(0,0)[r]{0}}
\put(1416,158){\rule[-0.175pt]{4.818pt}{0.350pt}}
\put(264,338){\rule[-0.175pt]{4.818pt}{0.350pt}}
\put(242,338){\makebox(0,0)[r]{1}}
\put(1416,338){\rule[-0.175pt]{4.818pt}{0.350pt}}
\put(264,518){\rule[-0.175pt]{4.818pt}{0.350pt}}
\put(242,518){\makebox(0,0)[r]{2}}
\put(1416,518){\rule[-0.175pt]{4.818pt}{0.350pt}}
\put(264,698){\rule[-0.175pt]{4.818pt}{0.350pt}}
\put(242,698){\makebox(0,0)[r]{3}}
\put(1416,698){\rule[-0.175pt]{4.818pt}{0.350pt}}
\put(264,877){\rule[-0.175pt]{4.818pt}{0.350pt}}
\put(242,877){\makebox(0,0)[r]{4}}
\put(1416,877){\rule[-0.175pt]{4.818pt}{0.350pt}}
\put(264,1057){\rule[-0.175pt]{4.818pt}{0.350pt}}
\put(242,1057){\makebox(0,0)[r]{5}}
\put(1416,1057){\rule[-0.175pt]{4.818pt}{0.350pt}}
\put(264,1237){\rule[-0.175pt]{4.818pt}{0.350pt}}
\put(242,1237){\makebox(0,0)[r]{6}}
\put(1416,1237){\rule[-0.175pt]{4.818pt}{0.350pt}}
\put(264,158){\rule[-0.175pt]{0.350pt}{4.818pt}}
\put(264,113){\makebox(0,0){0}}
\put(264,1217){\rule[-0.175pt]{0.350pt}{4.818pt}}
\put(498,158){\rule[-0.175pt]{0.350pt}{4.818pt}}
\put(498,113){\makebox(0,0){2}}
\put(498,1217){\rule[-0.175pt]{0.350pt}{4.818pt}}
\put(733,158){\rule[-0.175pt]{0.350pt}{4.818pt}}
\put(733,113){\makebox(0,0){4}}
\put(733,1217){\rule[-0.175pt]{0.350pt}{4.818pt}}
\put(967,158){\rule[-0.175pt]{0.350pt}{4.818pt}}
\put(967,113){\makebox(0,0){6}}
\put(967,1217){\rule[-0.175pt]{0.350pt}{4.818pt}}
\put(1202,158){\rule[-0.175pt]{0.350pt}{4.818pt}}
\put(1202,113){\makebox(0,0){8}}
\put(1202,1217){\rule[-0.175pt]{0.350pt}{4.818pt}}
\put(1436,158){\rule[-0.175pt]{0.350pt}{4.818pt}}
\put(1436,113){\makebox(0,0){10}}
\put(1436,1217){\rule[-0.175pt]{0.350pt}{4.818pt}}
\put(264,158){\rule[-0.175pt]{282.335pt}{0.350pt}}
\put(1436,158){\rule[-0.175pt]{0.350pt}{259.931pt}}
\put(264,1237){\rule[-0.175pt]{282.335pt}{0.350pt}}
\put(45,697){\makebox(0,0)[l]{\shortstack{$F(p)$}}}
\put(850,68){\makebox(0,0){$p^2$}}
\put(264,158){\rule[-0.175pt]{0.350pt}{259.931pt}}
\put(1306,1172){\makebox(0,0)[r]{t = 0.0}}
\put(1328,1172){\rule[-0.175pt]{15.899pt}{0.350pt}}
\put(265,178){\usebox{\plotpoint}}
\put(265,178){\rule[-0.175pt]{0.350pt}{1.124pt}}
\put(266,182){\rule[-0.175pt]{0.350pt}{1.124pt}}
\put(267,187){\rule[-0.175pt]{0.350pt}{1.124pt}}
\put(268,192){\rule[-0.175pt]{0.350pt}{0.803pt}}
\put(269,195){\rule[-0.175pt]{0.350pt}{0.803pt}}
\put(270,198){\rule[-0.175pt]{0.350pt}{0.803pt}}
\put(271,201){\rule[-0.175pt]{0.350pt}{0.562pt}}
\put(272,204){\rule[-0.175pt]{0.350pt}{0.562pt}}
\put(273,206){\rule[-0.175pt]{0.350pt}{0.562pt}}
\put(274,208){\rule[-0.175pt]{0.350pt}{0.482pt}}
\put(275,211){\rule[-0.175pt]{0.350pt}{0.482pt}}
\put(276,213){\rule[-0.175pt]{0.350pt}{0.482pt}}
\put(277,215){\rule[-0.175pt]{0.350pt}{0.482pt}}
\put(278,217){\rule[-0.175pt]{0.350pt}{0.482pt}}
\put(279,219){\rule[-0.175pt]{0.350pt}{0.482pt}}
\put(280,221){\usebox{\plotpoint}}
\put(281,222){\usebox{\plotpoint}}
\put(282,223){\usebox{\plotpoint}}
\put(283,224){\usebox{\plotpoint}}
\put(284,226){\usebox{\plotpoint}}
\put(285,227){\usebox{\plotpoint}}
\put(286,228){\usebox{\plotpoint}}
\put(287,230){\usebox{\plotpoint}}
\put(288,231){\usebox{\plotpoint}}
\put(289,232){\usebox{\plotpoint}}
\put(290,234){\usebox{\plotpoint}}
\put(291,235){\usebox{\plotpoint}}
\put(292,236){\usebox{\plotpoint}}
\put(292,237){\usebox{\plotpoint}}
\put(293,238){\usebox{\plotpoint}}
\put(294,239){\usebox{\plotpoint}}
\put(295,240){\usebox{\plotpoint}}
\put(296,241){\usebox{\plotpoint}}
\put(297,242){\usebox{\plotpoint}}
\put(298,243){\usebox{\plotpoint}}
\put(299,244){\usebox{\plotpoint}}
\put(300,245){\usebox{\plotpoint}}
\put(301,246){\rule[-0.175pt]{0.361pt}{0.350pt}}
\put(302,247){\rule[-0.175pt]{0.361pt}{0.350pt}}
\put(304,248){\rule[-0.175pt]{0.350pt}{0.361pt}}
\put(305,249){\rule[-0.175pt]{0.350pt}{0.361pt}}
\put(306,251){\rule[-0.175pt]{0.361pt}{0.350pt}}
\put(307,252){\rule[-0.175pt]{0.361pt}{0.350pt}}
\put(309,253){\rule[-0.175pt]{0.361pt}{0.350pt}}
\put(310,254){\rule[-0.175pt]{0.361pt}{0.350pt}}
\put(312,255){\rule[-0.175pt]{0.361pt}{0.350pt}}
\put(313,256){\rule[-0.175pt]{0.361pt}{0.350pt}}
\put(315,257){\rule[-0.175pt]{0.361pt}{0.350pt}}
\put(316,258){\rule[-0.175pt]{0.361pt}{0.350pt}}
\put(318,259){\rule[-0.175pt]{0.361pt}{0.350pt}}
\put(319,260){\rule[-0.175pt]{0.361pt}{0.350pt}}
\put(321,261){\rule[-0.175pt]{0.361pt}{0.350pt}}
\put(322,262){\rule[-0.175pt]{0.361pt}{0.350pt}}
\put(324,263){\rule[-0.175pt]{0.723pt}{0.350pt}}
\put(327,264){\rule[-0.175pt]{0.361pt}{0.350pt}}
\put(328,265){\rule[-0.175pt]{0.361pt}{0.350pt}}
\put(330,266){\rule[-0.175pt]{0.723pt}{0.350pt}}
\put(333,267){\rule[-0.175pt]{0.361pt}{0.350pt}}
\put(334,268){\rule[-0.175pt]{0.361pt}{0.350pt}}
\put(336,269){\rule[-0.175pt]{0.723pt}{0.350pt}}
\put(339,270){\rule[-0.175pt]{0.361pt}{0.350pt}}
\put(340,271){\rule[-0.175pt]{0.361pt}{0.350pt}}
\put(342,272){\rule[-0.175pt]{0.723pt}{0.350pt}}
\put(345,273){\rule[-0.175pt]{0.723pt}{0.350pt}}
\put(348,274){\rule[-0.175pt]{0.482pt}{0.350pt}}
\put(350,275){\rule[-0.175pt]{0.723pt}{0.350pt}}
\put(353,276){\rule[-0.175pt]{0.723pt}{0.350pt}}
\put(356,277){\rule[-0.175pt]{0.723pt}{0.350pt}}
\put(359,278){\rule[-0.175pt]{0.723pt}{0.350pt}}
\put(362,279){\rule[-0.175pt]{0.723pt}{0.350pt}}
\put(365,280){\rule[-0.175pt]{0.723pt}{0.350pt}}
\put(368,281){\rule[-0.175pt]{0.723pt}{0.350pt}}
\put(371,282){\rule[-0.175pt]{0.723pt}{0.350pt}}
\put(374,283){\rule[-0.175pt]{0.723pt}{0.350pt}}
\put(377,284){\rule[-0.175pt]{0.723pt}{0.350pt}}
\put(380,285){\rule[-0.175pt]{0.723pt}{0.350pt}}
\put(383,286){\rule[-0.175pt]{1.445pt}{0.350pt}}
\put(389,287){\rule[-0.175pt]{0.482pt}{0.350pt}}
\put(391,288){\rule[-0.175pt]{0.723pt}{0.350pt}}
\put(394,289){\rule[-0.175pt]{1.445pt}{0.350pt}}
\put(400,290){\rule[-0.175pt]{1.445pt}{0.350pt}}
\put(406,291){\rule[-0.175pt]{0.723pt}{0.350pt}}
\put(409,292){\rule[-0.175pt]{1.445pt}{0.350pt}}
\put(415,293){\rule[-0.175pt]{1.445pt}{0.350pt}}
\put(421,294){\rule[-0.175pt]{0.723pt}{0.350pt}}
\put(424,295){\rule[-0.175pt]{1.445pt}{0.350pt}}
\put(430,296){\rule[-0.175pt]{1.204pt}{0.350pt}}
\put(435,297){\rule[-0.175pt]{1.445pt}{0.350pt}}
\put(441,298){\rule[-0.175pt]{2.168pt}{0.350pt}}
\put(450,299){\rule[-0.175pt]{1.445pt}{0.350pt}}
\put(456,300){\rule[-0.175pt]{1.445pt}{0.350pt}}
\put(462,301){\rule[-0.175pt]{2.168pt}{0.350pt}}
\put(471,302){\rule[-0.175pt]{1.927pt}{0.350pt}}
\put(479,303){\rule[-0.175pt]{2.168pt}{0.350pt}}
\put(488,304){\rule[-0.175pt]{2.168pt}{0.350pt}}
\put(497,305){\rule[-0.175pt]{2.168pt}{0.350pt}}
\put(506,306){\rule[-0.175pt]{2.650pt}{0.350pt}}
\put(517,307){\rule[-0.175pt]{2.168pt}{0.350pt}}
\put(526,308){\rule[-0.175pt]{3.613pt}{0.350pt}}
\put(541,309){\rule[-0.175pt]{2.891pt}{0.350pt}}
\put(553,310){\rule[-0.175pt]{2.650pt}{0.350pt}}
\put(564,311){\rule[-0.175pt]{3.613pt}{0.350pt}}
\put(579,312){\rule[-0.175pt]{4.336pt}{0.350pt}}
\put(597,313){\rule[-0.175pt]{4.095pt}{0.350pt}}
\put(614,314){\rule[-0.175pt]{4.336pt}{0.350pt}}
\put(632,315){\rule[-0.175pt]{4.818pt}{0.350pt}}
\put(652,316){\rule[-0.175pt]{5.059pt}{0.350pt}}
\put(673,317){\rule[-0.175pt]{5.541pt}{0.350pt}}
\put(696,318){\rule[-0.175pt]{6.986pt}{0.350pt}}
\put(725,319){\rule[-0.175pt]{7.227pt}{0.350pt}}
\put(755,320){\rule[-0.175pt]{7.709pt}{0.350pt}}
\put(787,321){\rule[-0.175pt]{9.154pt}{0.350pt}}
\put(825,322){\rule[-0.175pt]{9.877pt}{0.350pt}}
\put(866,323){\rule[-0.175pt]{11.322pt}{0.350pt}}
\put(913,324){\rule[-0.175pt]{13.490pt}{0.350pt}}
\put(969,325){\rule[-0.175pt]{15.418pt}{0.350pt}}
\put(1033,326){\rule[-0.175pt]{18.308pt}{0.350pt}}
\put(1109,327){\rule[-0.175pt]{21.199pt}{0.350pt}}
\put(1197,328){\rule[-0.175pt]{26.981pt}{0.350pt}}
\put(1309,329){\rule[-0.175pt]{30.594pt}{0.350pt}}
\sbox{\plotpoint}{\rule[-0.250pt]{0.500pt}{0.500pt}}%
\put(1306,1127){\makebox(0,0)[r]{0.6}}
\put(1328,1127){\usebox{\plotpoint}}
\put(1348,1127){\usebox{\plotpoint}}
\put(1369,1127){\usebox{\plotpoint}}
\put(1390,1127){\usebox{\plotpoint}}
\put(1394,1127){\usebox{\plotpoint}}
\put(265,194){\usebox{\plotpoint}}
\put(265,194){\usebox{\plotpoint}}
\put(267,214){\usebox{\plotpoint}}
\put(270,235){\usebox{\plotpoint}}
\put(275,255){\usebox{\plotpoint}}
\put(282,274){\usebox{\plotpoint}}
\put(290,293){\usebox{\plotpoint}}
\put(300,311){\usebox{\plotpoint}}
\put(312,328){\usebox{\plotpoint}}
\put(327,343){\usebox{\plotpoint}}
\put(344,355){\usebox{\plotpoint}}
\put(362,366){\usebox{\plotpoint}}
\put(381,373){\usebox{\plotpoint}}
\put(400,380){\usebox{\plotpoint}}
\put(420,384){\usebox{\plotpoint}}
\put(440,388){\usebox{\plotpoint}}
\put(461,391){\usebox{\plotpoint}}
\put(481,393){\usebox{\plotpoint}}
\put(502,395){\usebox{\plotpoint}}
\put(522,396){\usebox{\plotpoint}}
\put(543,397){\usebox{\plotpoint}}
\put(563,398){\usebox{\plotpoint}}
\put(584,398){\usebox{\plotpoint}}
\put(605,398){\usebox{\plotpoint}}
\put(626,398){\usebox{\plotpoint}}
\put(646,397){\usebox{\plotpoint}}
\put(667,397){\usebox{\plotpoint}}
\put(688,396){\usebox{\plotpoint}}
\put(708,396){\usebox{\plotpoint}}
\put(729,395){\usebox{\plotpoint}}
\put(750,394){\usebox{\plotpoint}}
\put(770,393){\usebox{\plotpoint}}
\put(791,392){\usebox{\plotpoint}}
\put(812,392){\usebox{\plotpoint}}
\put(832,391){\usebox{\plotpoint}}
\put(853,390){\usebox{\plotpoint}}
\put(873,389){\usebox{\plotpoint}}
\put(894,388){\usebox{\plotpoint}}
\put(915,387){\usebox{\plotpoint}}
\put(935,386){\usebox{\plotpoint}}
\put(956,385){\usebox{\plotpoint}}
\put(976,384){\usebox{\plotpoint}}
\put(997,383){\usebox{\plotpoint}}
\put(1017,382){\usebox{\plotpoint}}
\put(1038,380){\usebox{\plotpoint}}
\put(1058,379){\usebox{\plotpoint}}
\put(1079,378){\usebox{\plotpoint}}
\put(1100,377){\usebox{\plotpoint}}
\put(1120,376){\usebox{\plotpoint}}
\put(1141,375){\usebox{\plotpoint}}
\put(1162,374){\usebox{\plotpoint}}
\put(1182,374){\usebox{\plotpoint}}
\put(1203,372){\usebox{\plotpoint}}
\put(1223,372){\usebox{\plotpoint}}
\put(1244,371){\usebox{\plotpoint}}
\put(1264,370){\usebox{\plotpoint}}
\put(1285,369){\usebox{\plotpoint}}
\put(1306,368){\usebox{\plotpoint}}
\put(1326,367){\usebox{\plotpoint}}
\put(1347,366){\usebox{\plotpoint}}
\put(1367,365){\usebox{\plotpoint}}
\put(1388,364){\usebox{\plotpoint}}
\put(1409,364){\usebox{\plotpoint}}
\put(1429,363){\usebox{\plotpoint}}
\put(1436,363){\usebox{\plotpoint}}
\put(1306,1082){\makebox(0,0)[r]{1.2}}
\put(1328,1082){\usebox{\plotpoint}}
\put(1365,1082){\usebox{\plotpoint}}
\put(1394,1082){\usebox{\plotpoint}}
\put(265,262){\usebox{\plotpoint}}
\put(265,262){\usebox{\plotpoint}}
\put(266,299){\usebox{\plotpoint}}
\put(268,336){\usebox{\plotpoint}}
\put(270,373){\usebox{\plotpoint}}
\put(273,411){\usebox{\plotpoint}}
\put(277,448){\usebox{\plotpoint}}
\put(282,485){\usebox{\plotpoint}}
\put(289,522){\usebox{\plotpoint}}
\put(298,558){\usebox{\plotpoint}}
\put(311,593){\usebox{\plotpoint}}
\put(336,619){\usebox{\plotpoint}}
\put(372,620){\usebox{\plotpoint}}
\put(405,603){\usebox{\plotpoint}}
\put(435,580){\usebox{\plotpoint}}
\put(464,557){\usebox{\plotpoint}}
\put(492,533){\usebox{\plotpoint}}
\put(521,510){\usebox{\plotpoint}}
\put(550,487){\usebox{\plotpoint}}
\put(580,465){\usebox{\plotpoint}}
\put(611,445){\usebox{\plotpoint}}
\put(643,425){\usebox{\plotpoint}}
\put(676,408){\usebox{\plotpoint}}
\put(709,392){\usebox{\plotpoint}}
\put(744,378){\usebox{\plotpoint}}
\put(779,366){\usebox{\plotpoint}}
\put(815,356){\usebox{\plotpoint}}
\put(851,347){\usebox{\plotpoint}}
\put(887,341){\usebox{\plotpoint}}
\put(923,335){\usebox{\plotpoint}}
\put(960,331){\usebox{\plotpoint}}
\put(997,328){\usebox{\plotpoint}}
\put(1034,326){\usebox{\plotpoint}}
\put(1071,325){\usebox{\plotpoint}}
\put(1108,325){\usebox{\plotpoint}}
\put(1146,325){\usebox{\plotpoint}}
\put(1183,326){\usebox{\plotpoint}}
\put(1220,327){\usebox{\plotpoint}}
\put(1257,329){\usebox{\plotpoint}}
\put(1294,331){\usebox{\plotpoint}}
\put(1331,332){\usebox{\plotpoint}}
\put(1368,334){\usebox{\plotpoint}}
\put(1405,337){\usebox{\plotpoint}}
\put(1436,338){\usebox{\plotpoint}}
\sbox{\plotpoint}{\rule[-0.500pt]{1.000pt}{1.000pt}}%
\put(1306,1037){\makebox(0,0)[r]{1.8}}
\put(1328,1037){\rule[-0.500pt]{15.899pt}{1.000pt}}
\put(265,470){\usebox{\plotpoint}}
\put(265,470){\rule[-0.500pt]{1.000pt}{16.943pt}}
\put(266,540){\rule[-0.500pt]{1.000pt}{16.943pt}}
\put(267,610){\rule[-0.500pt]{1.000pt}{16.943pt}}
\put(268,680){\rule[-0.500pt]{1.000pt}{10.359pt}}
\put(269,724){\rule[-0.500pt]{1.000pt}{10.359pt}}
\put(270,767){\rule[-0.500pt]{1.000pt}{10.359pt}}
\put(271,810){\rule[-0.500pt]{1.000pt}{7.468pt}}
\put(272,841){\rule[-0.500pt]{1.000pt}{7.468pt}}
\put(273,872){\rule[-0.500pt]{1.000pt}{7.468pt}}
\put(274,903){\rule[-0.500pt]{1.000pt}{5.782pt}}
\put(275,927){\rule[-0.500pt]{1.000pt}{5.782pt}}
\put(276,951){\rule[-0.500pt]{1.000pt}{5.782pt}}
\put(277,975){\rule[-0.500pt]{1.000pt}{4.497pt}}
\put(278,993){\rule[-0.500pt]{1.000pt}{4.497pt}}
\put(279,1012){\rule[-0.500pt]{1.000pt}{4.497pt}}
\put(280,1031){\rule[-0.500pt]{1.000pt}{3.613pt}}
\put(281,1046){\rule[-0.500pt]{1.000pt}{3.613pt}}
\put(282,1061){\rule[-0.500pt]{1.000pt}{3.613pt}}
\put(283,1076){\rule[-0.500pt]{1.000pt}{2.810pt}}
\put(284,1087){\rule[-0.500pt]{1.000pt}{2.810pt}}
\put(285,1099){\rule[-0.500pt]{1.000pt}{2.810pt}}
\put(286,1110){\rule[-0.500pt]{1.000pt}{2.248pt}}
\put(287,1120){\rule[-0.500pt]{1.000pt}{2.248pt}}
\put(288,1129){\rule[-0.500pt]{1.000pt}{2.248pt}}
\put(289,1139){\rule[-0.500pt]{1.000pt}{1.847pt}}
\put(290,1146){\rule[-0.500pt]{1.000pt}{1.847pt}}
\put(291,1154){\rule[-0.500pt]{1.000pt}{1.847pt}}
\put(292,1161){\rule[-0.500pt]{1.000pt}{1.365pt}}
\put(293,1167){\rule[-0.500pt]{1.000pt}{1.365pt}}
\put(294,1173){\rule[-0.500pt]{1.000pt}{1.365pt}}
\put(295,1178){\rule[-0.500pt]{1.000pt}{1.044pt}}
\put(296,1183){\rule[-0.500pt]{1.000pt}{1.044pt}}
\put(297,1187){\rule[-0.500pt]{1.000pt}{1.044pt}}
\put(298,1192){\usebox{\plotpoint}}
\put(299,1195){\usebox{\plotpoint}}
\put(300,1198){\usebox{\plotpoint}}
\put(301,1201){\usebox{\plotpoint}}
\put(302,1203){\usebox{\plotpoint}}
\put(303,1205){\usebox{\plotpoint}}
\put(304,1207){\usebox{\plotpoint}}
\put(305,1209){\usebox{\plotpoint}}
\put(306,1211){\usebox{\plotpoint}}
\put(309,1212){\usebox{\plotpoint}}
\put(312,1211){\usebox{\plotpoint}}
\put(313,1210){\usebox{\plotpoint}}
\put(314,1209){\usebox{\plotpoint}}
\put(315,1206){\usebox{\plotpoint}}
\put(316,1204){\usebox{\plotpoint}}
\put(317,1203){\usebox{\plotpoint}}
\put(318,1201){\usebox{\plotpoint}}
\put(319,1199){\usebox{\plotpoint}}
\put(320,1197){\usebox{\plotpoint}}
\put(321,1194){\usebox{\plotpoint}}
\put(322,1192){\usebox{\plotpoint}}
\put(323,1190){\usebox{\plotpoint}}
\put(324,1187){\usebox{\plotpoint}}
\put(325,1184){\usebox{\plotpoint}}
\put(326,1182){\usebox{\plotpoint}}
\put(327,1179){\usebox{\plotpoint}}
\put(328,1176){\usebox{\plotpoint}}
\put(329,1173){\usebox{\plotpoint}}
\put(330,1169){\usebox{\plotpoint}}
\put(331,1166){\usebox{\plotpoint}}
\put(332,1163){\usebox{\plotpoint}}
\put(333,1159){\usebox{\plotpoint}}
\put(334,1155){\usebox{\plotpoint}}
\put(335,1152){\usebox{\plotpoint}}
\put(336,1148){\usebox{\plotpoint}}
\put(337,1144){\usebox{\plotpoint}}
\put(338,1140){\usebox{\plotpoint}}
\put(339,1136){\usebox{\plotpoint}}
\put(340,1132){\usebox{\plotpoint}}
\put(341,1128){\usebox{\plotpoint}}
\put(342,1124){\usebox{\plotpoint}}
\put(343,1120){\usebox{\plotpoint}}
\put(344,1116){\usebox{\plotpoint}}
\put(345,1111){\rule[-0.500pt]{1.000pt}{1.044pt}}
\put(346,1107){\rule[-0.500pt]{1.000pt}{1.044pt}}
\put(347,1103){\rule[-0.500pt]{1.000pt}{1.044pt}}
\put(348,1096){\rule[-0.500pt]{1.000pt}{1.566pt}}
\put(349,1090){\rule[-0.500pt]{1.000pt}{1.566pt}}
\put(350,1085){\rule[-0.500pt]{1.000pt}{1.124pt}}
\put(351,1080){\rule[-0.500pt]{1.000pt}{1.124pt}}
\put(352,1076){\rule[-0.500pt]{1.000pt}{1.124pt}}
\put(353,1071){\rule[-0.500pt]{1.000pt}{1.044pt}}
\put(354,1067){\rule[-0.500pt]{1.000pt}{1.044pt}}
\put(355,1063){\rule[-0.500pt]{1.000pt}{1.044pt}}
\put(356,1058){\rule[-0.500pt]{1.000pt}{1.124pt}}
\put(357,1053){\rule[-0.500pt]{1.000pt}{1.124pt}}
\put(358,1049){\rule[-0.500pt]{1.000pt}{1.124pt}}
\put(359,1044){\rule[-0.500pt]{1.000pt}{1.205pt}}
\put(360,1039){\rule[-0.500pt]{1.000pt}{1.204pt}}
\put(361,1034){\rule[-0.500pt]{1.000pt}{1.204pt}}
\put(362,1029){\rule[-0.500pt]{1.000pt}{1.124pt}}
\put(363,1024){\rule[-0.500pt]{1.000pt}{1.124pt}}
\put(364,1020){\rule[-0.500pt]{1.000pt}{1.124pt}}
\put(365,1015){\rule[-0.500pt]{1.000pt}{1.124pt}}
\put(366,1010){\rule[-0.500pt]{1.000pt}{1.124pt}}
\put(367,1006){\rule[-0.500pt]{1.000pt}{1.124pt}}
\put(368,1001){\rule[-0.500pt]{1.000pt}{1.204pt}}
\put(369,996){\rule[-0.500pt]{1.000pt}{1.204pt}}
\put(370,991){\rule[-0.500pt]{1.000pt}{1.204pt}}
\put(371,986){\rule[-0.500pt]{1.000pt}{1.124pt}}
\put(372,981){\rule[-0.500pt]{1.000pt}{1.124pt}}
\put(373,977){\rule[-0.500pt]{1.000pt}{1.124pt}}
\put(374,972){\rule[-0.500pt]{1.000pt}{1.124pt}}
\put(375,967){\rule[-0.500pt]{1.000pt}{1.124pt}}
\put(376,963){\rule[-0.500pt]{1.000pt}{1.124pt}}
\put(377,958){\rule[-0.500pt]{1.000pt}{1.204pt}}
\put(378,953){\rule[-0.500pt]{1.000pt}{1.204pt}}
\put(379,948){\rule[-0.500pt]{1.000pt}{1.204pt}}
\put(380,943){\rule[-0.500pt]{1.000pt}{1.124pt}}
\put(381,938){\rule[-0.500pt]{1.000pt}{1.124pt}}
\put(382,934){\rule[-0.500pt]{1.000pt}{1.124pt}}
\put(383,929){\rule[-0.500pt]{1.000pt}{1.124pt}}
\put(384,924){\rule[-0.500pt]{1.000pt}{1.124pt}}
\put(385,920){\rule[-0.500pt]{1.000pt}{1.124pt}}
\put(386,915){\rule[-0.500pt]{1.000pt}{1.204pt}}
\put(387,910){\rule[-0.500pt]{1.000pt}{1.204pt}}
\put(388,905){\rule[-0.500pt]{1.000pt}{1.204pt}}
\put(389,898){\rule[-0.500pt]{1.000pt}{1.686pt}}
\put(390,891){\rule[-0.500pt]{1.000pt}{1.686pt}}
\put(391,886){\rule[-0.500pt]{1.000pt}{1.124pt}}
\put(392,881){\rule[-0.500pt]{1.000pt}{1.124pt}}
\put(393,877){\rule[-0.500pt]{1.000pt}{1.124pt}}
\put(394,872){\rule[-0.500pt]{1.000pt}{1.124pt}}
\put(395,867){\rule[-0.500pt]{1.000pt}{1.124pt}}
\put(396,863){\rule[-0.500pt]{1.000pt}{1.124pt}}
\put(397,858){\rule[-0.500pt]{1.000pt}{1.044pt}}
\put(398,854){\rule[-0.500pt]{1.000pt}{1.044pt}}
\put(399,850){\rule[-0.500pt]{1.000pt}{1.044pt}}
\put(400,845){\rule[-0.500pt]{1.000pt}{1.124pt}}
\put(401,840){\rule[-0.500pt]{1.000pt}{1.124pt}}
\put(402,836){\rule[-0.500pt]{1.000pt}{1.124pt}}
\put(403,831){\rule[-0.500pt]{1.000pt}{1.044pt}}
\put(404,827){\rule[-0.500pt]{1.000pt}{1.044pt}}
\put(405,823){\rule[-0.500pt]{1.000pt}{1.044pt}}
\put(406,818){\rule[-0.500pt]{1.000pt}{1.124pt}}
\put(407,813){\rule[-0.500pt]{1.000pt}{1.124pt}}
\put(408,809){\rule[-0.500pt]{1.000pt}{1.124pt}}
\put(409,804){\rule[-0.500pt]{1.000pt}{1.044pt}}
\put(410,800){\rule[-0.500pt]{1.000pt}{1.044pt}}
\put(411,796){\rule[-0.500pt]{1.000pt}{1.044pt}}
\put(412,791){\rule[-0.500pt]{1.000pt}{1.044pt}}
\put(413,787){\rule[-0.500pt]{1.000pt}{1.044pt}}
\put(414,783){\rule[-0.500pt]{1.000pt}{1.044pt}}
\put(415,778){\rule[-0.500pt]{1.000pt}{1.044pt}}
\put(416,774){\rule[-0.500pt]{1.000pt}{1.044pt}}
\put(417,770){\rule[-0.500pt]{1.000pt}{1.044pt}}
\put(418,766){\usebox{\plotpoint}}
\put(419,762){\usebox{\plotpoint}}
\put(420,758){\usebox{\plotpoint}}
\put(421,753){\rule[-0.500pt]{1.000pt}{1.044pt}}
\put(422,749){\rule[-0.500pt]{1.000pt}{1.044pt}}
\put(423,745){\rule[-0.500pt]{1.000pt}{1.044pt}}
\put(424,741){\usebox{\plotpoint}}
\put(425,737){\usebox{\plotpoint}}
\put(426,733){\usebox{\plotpoint}}
\put(427,729){\usebox{\plotpoint}}
\put(428,725){\usebox{\plotpoint}}
\put(429,721){\usebox{\plotpoint}}
\put(430,715){\rule[-0.500pt]{1.000pt}{1.445pt}}
\put(431,709){\rule[-0.500pt]{1.000pt}{1.445pt}}
\put(432,705){\usebox{\plotpoint}}
\put(433,701){\usebox{\plotpoint}}
\put(434,698){\usebox{\plotpoint}}
\put(435,694){\usebox{\plotpoint}}
\put(436,690){\usebox{\plotpoint}}
\put(437,686){\usebox{\plotpoint}}
\put(438,682){\usebox{\plotpoint}}
\put(439,678){\usebox{\plotpoint}}
\put(440,675){\usebox{\plotpoint}}
\put(441,671){\usebox{\plotpoint}}
\put(442,667){\usebox{\plotpoint}}
\put(443,664){\usebox{\plotpoint}}
\put(444,660){\usebox{\plotpoint}}
\put(445,656){\usebox{\plotpoint}}
\put(446,653){\usebox{\plotpoint}}
\put(447,649){\usebox{\plotpoint}}
\put(448,646){\usebox{\plotpoint}}
\put(449,643){\usebox{\plotpoint}}
\put(450,639){\usebox{\plotpoint}}
\put(451,635){\usebox{\plotpoint}}
\put(452,632){\usebox{\plotpoint}}
\put(453,628){\usebox{\plotpoint}}
\put(454,625){\usebox{\plotpoint}}
\put(455,622){\usebox{\plotpoint}}
\put(456,618){\usebox{\plotpoint}}
\put(457,615){\usebox{\plotpoint}}
\put(458,612){\usebox{\plotpoint}}
\put(459,608){\usebox{\plotpoint}}
\put(460,605){\usebox{\plotpoint}}
\put(461,602){\usebox{\plotpoint}}
\put(462,599){\usebox{\plotpoint}}
\put(463,596){\usebox{\plotpoint}}
\put(464,593){\usebox{\plotpoint}}
\put(465,589){\usebox{\plotpoint}}
\put(466,586){\usebox{\plotpoint}}
\put(467,583){\usebox{\plotpoint}}
\put(468,580){\usebox{\plotpoint}}
\put(469,577){\usebox{\plotpoint}}
\put(470,574){\usebox{\plotpoint}}
\put(471,569){\rule[-0.500pt]{1.000pt}{1.084pt}}
\put(472,565){\rule[-0.500pt]{1.000pt}{1.084pt}}
\put(473,562){\usebox{\plotpoint}}
\put(474,559){\usebox{\plotpoint}}
\put(475,556){\usebox{\plotpoint}}
\put(476,553){\usebox{\plotpoint}}
\put(477,550){\usebox{\plotpoint}}
\put(478,548){\usebox{\plotpoint}}
\put(479,545){\usebox{\plotpoint}}
\put(480,542){\usebox{\plotpoint}}
\put(481,540){\usebox{\plotpoint}}
\put(482,537){\usebox{\plotpoint}}
\put(483,534){\usebox{\plotpoint}}
\put(484,531){\usebox{\plotpoint}}
\put(485,528){\usebox{\plotpoint}}
\put(486,526){\usebox{\plotpoint}}
\put(487,524){\usebox{\plotpoint}}
\put(488,521){\usebox{\plotpoint}}
\put(489,518){\usebox{\plotpoint}}
\put(490,516){\usebox{\plotpoint}}
\put(491,513){\usebox{\plotpoint}}
\put(492,510){\usebox{\plotpoint}}
\put(493,508){\usebox{\plotpoint}}
\put(494,505){\usebox{\plotpoint}}
\put(495,503){\usebox{\plotpoint}}
\put(496,501){\usebox{\plotpoint}}
\put(497,498){\usebox{\plotpoint}}
\put(498,496){\usebox{\plotpoint}}
\put(499,494){\usebox{\plotpoint}}
\put(500,491){\usebox{\plotpoint}}
\put(501,489){\usebox{\plotpoint}}
\put(502,487){\usebox{\plotpoint}}
\put(503,484){\usebox{\plotpoint}}
\put(504,482){\usebox{\plotpoint}}
\put(505,480){\usebox{\plotpoint}}
\put(506,477){\usebox{\plotpoint}}
\put(507,475){\usebox{\plotpoint}}
\put(508,473){\usebox{\plotpoint}}
\put(509,471){\usebox{\plotpoint}}
\put(510,469){\usebox{\plotpoint}}
\put(511,467){\usebox{\plotpoint}}
\put(512,465){\usebox{\plotpoint}}
\put(513,463){\usebox{\plotpoint}}
\put(514,461){\usebox{\plotpoint}}
\put(515,457){\usebox{\plotpoint}}
\put(516,454){\usebox{\plotpoint}}
\put(517,452){\usebox{\plotpoint}}
\put(518,450){\usebox{\plotpoint}}
\put(519,449){\usebox{\plotpoint}}
\put(520,447){\usebox{\plotpoint}}
\put(521,445){\usebox{\plotpoint}}
\put(522,443){\usebox{\plotpoint}}
\put(523,441){\usebox{\plotpoint}}
\put(524,439){\usebox{\plotpoint}}
\put(525,437){\usebox{\plotpoint}}
\put(526,435){\usebox{\plotpoint}}
\put(527,433){\usebox{\plotpoint}}
\put(528,432){\usebox{\plotpoint}}
\put(529,430){\usebox{\plotpoint}}
\put(530,428){\usebox{\plotpoint}}
\put(531,426){\usebox{\plotpoint}}
\put(532,424){\usebox{\plotpoint}}
\put(533,422){\usebox{\plotpoint}}
\put(534,421){\usebox{\plotpoint}}
\put(535,419){\usebox{\plotpoint}}
\put(536,417){\usebox{\plotpoint}}
\put(537,416){\usebox{\plotpoint}}
\put(538,414){\usebox{\plotpoint}}
\put(539,413){\usebox{\plotpoint}}
\put(540,412){\usebox{\plotpoint}}
\put(541,410){\usebox{\plotpoint}}
\put(542,408){\usebox{\plotpoint}}
\put(543,407){\usebox{\plotpoint}}
\put(544,405){\usebox{\plotpoint}}
\put(545,403){\usebox{\plotpoint}}
\put(546,402){\usebox{\plotpoint}}
\put(547,400){\usebox{\plotpoint}}
\put(548,399){\usebox{\plotpoint}}
\put(549,398){\usebox{\plotpoint}}
\put(550,396){\usebox{\plotpoint}}
\put(551,395){\usebox{\plotpoint}}
\put(552,394){\usebox{\plotpoint}}
\put(553,392){\usebox{\plotpoint}}
\put(554,391){\usebox{\plotpoint}}
\put(555,390){\usebox{\plotpoint}}
\put(556,388){\usebox{\plotpoint}}
\put(557,386){\usebox{\plotpoint}}
\put(558,384){\usebox{\plotpoint}}
\put(559,383){\usebox{\plotpoint}}
\put(560,382){\usebox{\plotpoint}}
\put(561,380){\usebox{\plotpoint}}
\put(562,379){\usebox{\plotpoint}}
\put(563,378){\usebox{\plotpoint}}
\put(564,378){\usebox{\plotpoint}}
\put(565,377){\usebox{\plotpoint}}
\put(566,376){\usebox{\plotpoint}}
\put(567,373){\usebox{\plotpoint}}
\put(568,372){\usebox{\plotpoint}}
\put(569,371){\usebox{\plotpoint}}
\put(570,371){\usebox{\plotpoint}}
\put(571,370){\usebox{\plotpoint}}
\put(572,369){\usebox{\plotpoint}}
\put(573,368){\usebox{\plotpoint}}
\put(574,367){\usebox{\plotpoint}}
\put(575,366){\usebox{\plotpoint}}
\put(576,365){\usebox{\plotpoint}}
\put(577,364){\usebox{\plotpoint}}
\put(578,363){\usebox{\plotpoint}}
\put(579,362){\usebox{\plotpoint}}
\put(580,361){\usebox{\plotpoint}}
\put(581,360){\usebox{\plotpoint}}
\put(582,359){\usebox{\plotpoint}}
\put(583,358){\usebox{\plotpoint}}
\put(584,357){\usebox{\plotpoint}}
\put(585,356){\usebox{\plotpoint}}
\put(586,355){\usebox{\plotpoint}}
\put(588,354){\usebox{\plotpoint}}
\put(589,353){\usebox{\plotpoint}}
\put(590,352){\usebox{\plotpoint}}
\put(591,351){\usebox{\plotpoint}}
\put(592,350){\usebox{\plotpoint}}
\put(594,349){\usebox{\plotpoint}}
\put(595,348){\usebox{\plotpoint}}
\put(596,347){\usebox{\plotpoint}}
\put(597,346){\usebox{\plotpoint}}
\put(598,345){\usebox{\plotpoint}}
\put(599,344){\usebox{\plotpoint}}
\put(600,343){\usebox{\plotpoint}}
\put(602,342){\usebox{\plotpoint}}
\put(603,341){\usebox{\plotpoint}}
\put(605,340){\usebox{\plotpoint}}
\put(606,339){\usebox{\plotpoint}}
\put(608,338){\usebox{\plotpoint}}
\put(609,337){\usebox{\plotpoint}}
\put(611,336){\usebox{\plotpoint}}
\put(612,335){\usebox{\plotpoint}}
\put(614,334){\usebox{\plotpoint}}
\put(615,333){\usebox{\plotpoint}}
\put(617,332){\usebox{\plotpoint}}
\put(620,331){\usebox{\plotpoint}}
\put(621,330){\usebox{\plotpoint}}
\put(623,329){\usebox{\plotpoint}}
\put(626,328){\usebox{\plotpoint}}
\put(627,327){\usebox{\plotpoint}}
\put(629,326){\usebox{\plotpoint}}
\put(632,325){\usebox{\plotpoint}}
\put(635,324){\usebox{\plotpoint}}
\put(638,323){\usebox{\plotpoint}}
\put(641,322){\usebox{\plotpoint}}
\put(643,321){\usebox{\plotpoint}}
\put(646,320){\usebox{\plotpoint}}
\put(649,319){\usebox{\plotpoint}}
\put(652,318){\rule[-0.500pt]{1.445pt}{1.000pt}}
\put(658,317){\usebox{\plotpoint}}
\put(661,316){\rule[-0.500pt]{1.445pt}{1.000pt}}
\put(667,315){\rule[-0.500pt]{1.445pt}{1.000pt}}
\put(673,314){\rule[-0.500pt]{2.650pt}{1.000pt}}
\put(684,313){\rule[-0.500pt]{5.782pt}{1.000pt}}
\put(708,314){\rule[-0.500pt]{2.891pt}{1.000pt}}
\put(720,315){\rule[-0.500pt]{1.927pt}{1.000pt}}
\put(728,316){\rule[-0.500pt]{2.168pt}{1.000pt}}
\put(737,317){\rule[-0.500pt]{1.445pt}{1.000pt}}
\put(743,318){\rule[-0.500pt]{1.445pt}{1.000pt}}
\put(749,319){\rule[-0.500pt]{1.445pt}{1.000pt}}
\put(755,320){\rule[-0.500pt]{1.445pt}{1.000pt}}
\put(761,321){\usebox{\plotpoint}}
\put(764,322){\rule[-0.500pt]{1.204pt}{1.000pt}}
\put(769,323){\usebox{\plotpoint}}
\put(772,324){\rule[-0.500pt]{1.445pt}{1.000pt}}
\put(778,325){\usebox{\plotpoint}}
\put(781,326){\rule[-0.500pt]{1.445pt}{1.000pt}}
\put(787,327){\usebox{\plotpoint}}
\put(790,328){\rule[-0.500pt]{1.445pt}{1.000pt}}
\put(796,329){\usebox{\plotpoint}}
\put(799,330){\usebox{\plotpoint}}
\put(802,331){\rule[-0.500pt]{1.445pt}{1.000pt}}
\put(808,332){\usebox{\plotpoint}}
\put(810,333){\usebox{\plotpoint}}
\put(813,334){\rule[-0.500pt]{1.445pt}{1.000pt}}
\put(819,335){\usebox{\plotpoint}}
\put(822,336){\usebox{\plotpoint}}
\put(825,337){\rule[-0.500pt]{1.445pt}{1.000pt}}
\put(831,338){\usebox{\plotpoint}}
\put(834,339){\usebox{\plotpoint}}
\put(837,340){\rule[-0.500pt]{1.445pt}{1.000pt}}
\put(843,341){\usebox{\plotpoint}}
\put(846,342){\usebox{\plotpoint}}
\put(849,343){\rule[-0.500pt]{1.204pt}{1.000pt}}
\put(854,344){\usebox{\plotpoint}}
\put(857,345){\usebox{\plotpoint}}
\put(860,346){\rule[-0.500pt]{1.445pt}{1.000pt}}
\put(866,347){\usebox{\plotpoint}}
\put(869,348){\rule[-0.500pt]{1.445pt}{1.000pt}}
\put(875,349){\usebox{\plotpoint}}
\put(878,350){\usebox{\plotpoint}}
\put(881,351){\rule[-0.500pt]{1.445pt}{1.000pt}}
\put(887,352){\usebox{\plotpoint}}
\put(890,353){\rule[-0.500pt]{1.204pt}{1.000pt}}
\put(895,354){\usebox{\plotpoint}}
\put(898,355){\rule[-0.500pt]{1.445pt}{1.000pt}}
\put(904,356){\rule[-0.500pt]{1.445pt}{1.000pt}}
\put(910,357){\usebox{\plotpoint}}
\put(913,358){\rule[-0.500pt]{1.445pt}{1.000pt}}
\put(919,359){\rule[-0.500pt]{1.445pt}{1.000pt}}
\put(925,360){\usebox{\plotpoint}}
\put(928,361){\rule[-0.500pt]{1.445pt}{1.000pt}}
\put(934,362){\rule[-0.500pt]{1.204pt}{1.000pt}}
\put(939,363){\rule[-0.500pt]{1.445pt}{1.000pt}}
\put(945,364){\rule[-0.500pt]{1.445pt}{1.000pt}}
\put(951,365){\rule[-0.500pt]{1.445pt}{1.000pt}}
\put(957,366){\rule[-0.500pt]{2.168pt}{1.000pt}}
\put(966,367){\rule[-0.500pt]{1.445pt}{1.000pt}}
\put(972,368){\rule[-0.500pt]{1.927pt}{1.000pt}}
\put(980,369){\rule[-0.500pt]{2.168pt}{1.000pt}}
\put(989,370){\rule[-0.500pt]{2.168pt}{1.000pt}}
\put(998,371){\rule[-0.500pt]{2.168pt}{1.000pt}}
\put(1007,372){\rule[-0.500pt]{2.650pt}{1.000pt}}
\put(1018,373){\rule[-0.500pt]{3.613pt}{1.000pt}}
\put(1033,374){\rule[-0.500pt]{5.059pt}{1.000pt}}
\put(1054,375){\rule[-0.500pt]{16.863pt}{1.000pt}}
\put(1124,374){\rule[-0.500pt]{5.541pt}{1.000pt}}
\put(1147,373){\rule[-0.500pt]{4.336pt}{1.000pt}}
\put(1165,372){\rule[-0.500pt]{3.613pt}{1.000pt}}
\put(1180,371){\rule[-0.500pt]{2.650pt}{1.000pt}}
\put(1191,370){\rule[-0.500pt]{2.891pt}{1.000pt}}
\put(1203,369){\rule[-0.500pt]{2.891pt}{1.000pt}}
\put(1215,368){\rule[-0.500pt]{2.168pt}{1.000pt}}
\put(1224,367){\rule[-0.500pt]{2.650pt}{1.000pt}}
\put(1235,366){\rule[-0.500pt]{2.168pt}{1.000pt}}
\put(1244,365){\rule[-0.500pt]{2.891pt}{1.000pt}}
\put(1256,364){\rule[-0.500pt]{2.168pt}{1.000pt}}
\put(1265,363){\rule[-0.500pt]{1.927pt}{1.000pt}}
\put(1273,362){\rule[-0.500pt]{2.168pt}{1.000pt}}
\put(1282,361){\rule[-0.500pt]{2.168pt}{1.000pt}}
\put(1291,360){\rule[-0.500pt]{2.168pt}{1.000pt}}
\put(1300,359){\rule[-0.500pt]{2.168pt}{1.000pt}}
\put(1309,358){\rule[-0.500pt]{1.927pt}{1.000pt}}
\put(1317,357){\rule[-0.500pt]{2.168pt}{1.000pt}}
\put(1326,356){\rule[-0.500pt]{2.168pt}{1.000pt}}
\put(1335,355){\rule[-0.500pt]{2.168pt}{1.000pt}}
\put(1344,354){\rule[-0.500pt]{1.927pt}{1.000pt}}
\put(1352,353){\rule[-0.500pt]{2.168pt}{1.000pt}}
\put(1361,352){\rule[-0.500pt]{2.168pt}{1.000pt}}
\put(1370,351){\rule[-0.500pt]{2.168pt}{1.000pt}}
\put(1379,350){\rule[-0.500pt]{2.168pt}{1.000pt}}
\put(1388,349){\rule[-0.500pt]{1.927pt}{1.000pt}}
\put(1396,348){\rule[-0.500pt]{2.168pt}{1.000pt}}
\put(1405,347){\rule[-0.500pt]{2.168pt}{1.000pt}}
\put(1414,346){\rule[-0.500pt]{2.168pt}{1.000pt}}
\put(1423,345){\rule[-0.500pt]{2.891pt}{1.000pt}}
\put(1435,344){\usebox{\plotpoint}}
\end{picture}
}

\caption{The decay of the false vacuum. Initial mass $m = 1$,
$\lambda = 1.0$. The sign of the mass squared term is reversed at
$t=t_0=0$ by switching on $M^2 = 2$. (Units defined by
$c=\hbar=1$.)\label{fig1}}

\end{figure}

The non-trivial evolution starts at $t=t_0$ when we switch on the $-M^2$ term.
It was chosen large enough to override
the initial positive $m^2$ and the system rolls down  towards
the new ground state.
The results are shown in Fig.
\ref{fig1}. The maximum at low momenta indicates the
development of long-range correlation as shown by
Eq. (\ref{SbrkWF}). (It is not exactly a delta function because
we are dealing with finite times.) A more detailed analysis
shows that the `classical component' of the field reaches the
value given by (\ref{Bsol}) for large times when the oscillations
around the new minimum die out. Both the initial decay and damping
of oscillations are purely quantum processes (the latter results
from particle creation by a time-dependent scalar potential).
We should stress that this numerical analysis was intended only
as an illustration and there are still some questions left.
One of them is the stability of the procedure described here
although we observe the good numerical stability of our results.

The methods developed here allow for the study of far more
complicated transitions in presence of abelian gauge fields, spatial
inhomogeneities etc.

Our formalism does not include the temperature effects which are
of great importance for the phase transitions (e.g., in
cosmology).  This is an obvious shortcoming. We think that
thermal effects can be incorporated at moderate expense.

\section{Summary}                                          \label{secSumm}
Starting from the definition of the Wigner transform as a realization of the
Heisenberg algebra in phase space, we have developed a calculus to translate
operator equations into phase-space equations that have an expansion in
powers of $\hbar$ and clearly exhibit the classical limit. Applying this
formalism to the symmetrized correlation function of the Klein--Gordon field,
we have derived the equation of motion for the equal-time Wigner transform
of the Klein--Gordon field and given a meaning to its components. The
resulting set of equations has been shown to possess an interpretation in
terms of a relativistic gas of negatively and positively charged particles.
Interference terms gave rise to quantum corrections and, especially,
Schwinger pair creation. This formulation of the Klein--Gordon field is
especially suited to semiclassical nonperturbative calculations as is
demonstrated in the case of spontaneous symmetry breaking.
\\[10pt]
\noindent{\em Acknowledgements:}
P.G.~would like to acknowledge discussions with Professor Iwo
Bialynicki-Birula and correspondence from Professor Johann Rafelski.
C.B.~wishes to thank Judah M.~Eisenberg, Stefan Graf, Andreas Sch\"afer, and
Gerhard Soff for many discussions and careful reading of the manuscript, and
the German Israel Foundation for its support.

\clearpage
\section*{Figures}


\begin{thebibliography}{XXXX}
\bibitem{BGR} {\sc I. Bialynicki--Birula, P. G\' ornicki, and J.
Rafelski}, {\em Phys. Rev.} D {\bf 44}, 1825 (1991).
\bibitem{Wigner} {\sc E. P. Wigner}, {\em Phys. Rev.} {\bf 32}, 749 (1932).
\bibitem{Eisen} {\sc J. M. Eisenberg and G. K\" albermann}, {\em Phys. Rev.} D
{\bf 37}, 1197 (1988).
\bibitem{Vasak} {\sc D. Vasak, M. Gyulassy, and H.-Th. Elze},
{\em Ann. Phys. (N.Y.)} {\bf 173}, 462 (1987).
\bibitem{Elze} {\sc H.-Th. Elze and U. Heinz}, {\em Phys. Rep.} {\bf 183}, 81
(1989).
\bibitem{CM1} {\sc F. Cooper and E. Mottola}, {\em Phys. Rev.} D {\bf 36},
3114 (1987).
\bibitem{CM2} {\sc F. Cooper and E. Mottola}, {\em Phys. Rev.} D {\bf 40}, 456
(1989)
\bibitem{CM3} {\sc F. Cooper, E. Mottola, B. Rogers, and P. Anderson},
{\em in} ``Intermittency in High Energy Collisions'' (F. Cooper et
al., Eds.), p.~399, World Scientific, Singapore, 1991.
\bibitem{Provid} {\sc J. da Provid\^ encia, M. C. Ruivo, and C. A. de
Sousa}, {\em Phys. Rev.} D {\bf 36}, 1882 (1987).
\bibitem{Hil} {\sc M. Hillery, R. F. O'Connell, M. O. Scully, and E. P.
Wigner},
{\em Phys. Rep.} {\bf 106}, 121 (1984).
\bibitem{FV} {\sc H. Feshbach and F. Villars}, {\em Rev. Mod. Phys.} {\bf 30},
24 (1958).
\bibitem{Davydov} {\sc A. S. Davydov}, ``Quantum mechanics'' 2nd ed.,
Pergamon Press, Oxford, 1976.
\bibitem{Schw} {\sc J. Schwinger}, {\em Phys. Rev.} {\bf 82}, 664 (1951).
\bibitem{next} {\sc C. Best, J. M. Eisenberg},
``Pair creation in transport equations using the equal-time Wigner
transform'', Tel Aviv University Physics Preprint TAUP, September 1992.
\end{thebibliography}
\end{document}